\documentclass{SciPost}

\hypersetup{
    colorlinks,
    linkcolor={red!50!black},
    citecolor={blue!50!black},
    urlcolor={blue!80!black}
}

\usepackage[table]{xcolor}
\usepackage{array}
\usepackage{multirow}
\usepackage{xspace}
\usepackage{dirtree}
\usepackage{setspace}
\usepackage[bitstream-charter]{mathdesign}
\usepackage{soul}
 
\DeclareSymbolFont{usualmathcal}{OMS}{cmsy}{m}{n}
\DeclareSymbolFontAlphabet{\mathcal}{usualmathcal}

\newcommand{\matchete}{\texttt{Matchete}\xspace}
\newcommand{\OpToCpp}{\texttt{OperatorToC++}\xspace}

\fancypagestyle{SPstyle}{
\fancyhf{}
\lhead{\colorbox{scipostblue}{\bf \color{white} ~SciPost Physics Codebases }}
\rhead{{\bf \color{scipostdeepblue} ~Submission }}

\fancyfoot[C]{\textbf{\thepage}}
}

\usepackage{listings}
\usepackage{xcolor}

\definecolor{codebg}{rgb}{0.97,0.97,0.97}     
\definecolor{codekw}{rgb}{0,0,0.6}            
\definecolor{codestring}{rgb}{0.5,0,0}        
\definecolor{codecomment}{rgb}{0.25,0.5,0.35} 
\definecolor{codenumber}{rgb}{0.5,0.5,0.5}    
\definecolor{mathexin}{rgb}{0.2,0.4,0.8}
\definecolor{mathexout}{rgb}{0.6,0.7,0.8}
\definecolor{codenumber}{rgb}{0.5,0.5,0.5}

\lstdefinestyle{base}{
    backgroundcolor=\color{codebg},
    basicstyle=\ttfamily\fontsize{8}{10}\selectfont,
    breaklines=true,
    captionpos=b,
    frame=single,
    frameround=tttt,
    rulecolor=\color{black},
    numbers=left,
    numbersep=6pt,
    numberstyle=\tiny\color{codenumber},
    keywordstyle=\color{codekw}\bfseries,
    stringstyle=\color{codestring},
    commentstyle=\color{codecomment}\itshape,
    showstringspaces=false,
    tabsize=4
}

\lstdefinestyle{Python}{
    style=base,
    language=Python,
    morekeywords={self, as, with}
}

\lstdefinestyle{cpp}{
    style=base,
    language=C++,
    morekeywords={constexpr,nullptr,size_t}
}

\lstdefinelanguage{MathematicaPrompt}{
  morekeywords={Plot,Sin,Cos,Exp,Factorial,Table,Module,If,Print,Sqrt,Pi},
  sensitive=true,
  morecomment=[l]{(*},
  morecomment=[s]{(*}{*)},
  morestring=[b]"
}

\newcounter{mathex}
\newcommand{\InPromptNext}{%
  \stepcounter{mathex}%
  \llap{\makebox[3.5em][r]{\textcolor{mathexin}{\fontsize{7}{8}\selectfont\sffamily In[\themathex]:=}\hspace{0.1cm}}}%
}
\newcommand{\OutPromptOfCurrent}{%
  \llap{\makebox[3.5em][r]{\textcolor{mathexout}{\fontsize{7}{8}\selectfont\sffamily Out[\themathex]=}\hspace{0.1cm}}}%
}

\lstdefinestyle{mathematicaMargin}{
    style=base,
    language=MathematicaPrompt,
    numbers=none,
    xleftmargin=4.5em,
    framexleftmargin=3.8em,
    escapechar=?
}

\newcommand{\smallsquare}[1]{\textcolor{#1}{\rule{1.2ex}{1.2ex}}}

\definecolor{myTeal}{RGB}{0, 128, 128}
\colorlet{tealL1}{myTeal!20}
\colorlet{tealL2}{myTeal!60}
\colorlet{tealL3}{myTeal}

\usepackage{tikz}
\usetikzlibrary{shapes.geometric, arrows, positioning, fit, calc}

\tikzstyle{darkbox} = [rectangle, rounded corners, minimum width=3cm, minimum height=1cm, text centered, font=\small\sffamily, draw=black, line width=1.2pt, text=white, fill=tealL3]
\tikzstyle{standardbox} = [rectangle, rounded corners, minimum width=3cm, minimum height=1cm, text centered, font=\small\sffamily, draw=black, fill=tealL1]
\tikzstyle{lightbox} = [rectangle, rounded corners, minimum width=3cm, minimum height=1cm, text centered, font=\small\sffamily, draw=black, fill=tealL2]

\tikzstyle{arrow} = [thick, ->, >=stealth, shorten <=2pt, shorten >=2pt]
\tikzstyle{group_dash} = [rectangle, draw=black, dashed, inner sep=0.4cm]
\tikzstyle{group} = [rectangle, draw=black, inner sep=0.2cm]

\tikzstyle{folder} = [draw=none, inner sep=2pt]
\begin{document}

\pagestyle{SPstyle}

\begin{center}{\Large \textbf{\color{scipostdeepblue}{
OperatorToC++: Transpiling Matched EFT Coefficients \\to Low-Level Routines\\
}}}\end{center}

\begin{center}\textbf{
Sabine Kraml\textsuperscript{1$\star$},
Andre Lessa\textsuperscript{2$\dagger$},
Suraj Prakash\textsuperscript{3$\ddagger$} and
Felix Wilsch\textsuperscript{4$\S$}
}\end{center}

\begin{center}
{\bf 1} Laboratoire de Physique Subatomique et de Cosmologie, Université Grenoble-Alpes, \\CNRS/IN2P3, 53 Avenue des Martyrs, 38026 Grenoble, France
\\
{\bf 2} Instituto de F\'isica, Universidade de S\~ao Paulo, \\05315-970 SP, Brazil
\\
{\bf 3} Instituto de F\'isica Corpuscular (IFIC), CSIC-Universitat de Val\`encia (UV),\\ 
C\,/ Catedrático José Beltrán 2, E-46980 Paterna (Valencia), Spain
\\
{\bf 4} Institute for Theoretical Particle Physics and Cosmology, RWTH Aachen University, \\Sommerfeldstr.~16 , D-52074 Aachen, Germany
\\[\baselineskip]
$\star$ \href{mailto:sabine.kraml@lpsc.in2p3.fr}{\small sabine.kraml@lpsc.in2p3.fr}\,,\quad
$\dagger$ \href{mailto:lessa@if.usp.br}{\small lessa@if.usp.br}\,,\\
$\ddagger$ \href{mailto:suraj.prakash@ific.uv.es}{\small suraj.prakash@ific.uv.es}\,,\quad
$\S$ \href{mailto:felix.wilsch@physik.rwth-aachen.de}{\small felix.wilsch@physik.rwth-aachen.de}.
\end{center}

\section*{\color{scipostdeepblue}{Abstract}}
\textbf{\boldmath{%
In recent years, significant progress has been made in the development of automated tools that match the parameters of new physics models and the appropriate low-energy Effective Field Theories. This work introduces an extensible, hybrid tool, \OpToCpp, that combines the strengths of Mathematica and C++ to facilitate the next steps beyond the matching. \OpToCpp efficiently tackles the complexities within the ana\-lyti\-cal matched expressions such as intricate loop-functions and lengthy sums and products involving tensor objects. It then translates and bundles the results into C++ classes and functions which provide a convenient platform for further numerical analyses. Finally, it offers the possibility of calling the compiled Wilson coefficient methods as Python functions, thus enabling to link them with the vast Python library ecosystem for High Energy Physics workflows.
}}

\vspace{\baselineskip}



\noindent\rule{\textwidth}{1pt}
\tableofcontents
\noindent\rule{\textwidth}{1pt}

\section{Introduction}

The first quarter of the twenty-first century witnessed monumental progress in the field of high energy physics. While the first observation of the Higgs boson was a major landmark~\cite{ATLAS:2012yve,CMS:2012qbp}, further progress in experimental techniques made it possible to detect low-energy neutrino-nucleus scattering~\cite{COHERENT:2017ipa} and (ultra-)high energy signals from transient astrophysical sources~\cite{KM3NeT:2025npi} for the first time. Large datasets from low- as well as high-energy experiments continue to corroborate the predictions of the Standard Model~(SM) with increasing precision. Yet, several features of the universe, e.g. neutrino masses, dark matter, matter-antimatter asymmetry, remain unaccounted for by the SM. Explaining
such phenomena requires improved theoretical modelling and a meticulous scrutiny of such models in light of the data obtained for distinct processes across a diverse energy range.

This is where the framework of Effective Field Theory~(EFT)~\cite{Buchmuller:1985jz} has been found to be of great utility, especially to conduct model-independent analyses of beyond-the-SM effects. EFT-based studies follow one of two complimentary approaches. The \textit{bottom-up} approach involves the construction of a complete set of operators, that describe contact interactions among the low-energy degrees of freedom, guided by the symmetries of the theory. Individual operators are accompanied by free parameters, known as Wilson coefficients, and organised based on their mass dimensions. The \textit{top-down} approach involves suitably identifying and integrating out heavy degrees of freedom and matching the parameters of a high-scale model with the corresponding low-energy EFT. 
In recent years, a number of computational tools have been developed to automate and simplify the lengthy calculations involved in the matching of parameters, e.g.\ \matchete~\cite{Fuentes-Martin:2022jrf}, \texttt{MatchmakerEFT}~\cite{Carmona:2021xtq}, and \texttt{CoDEx}~\cite{DasBakshi:2018vni}.

In a recent project \cite{Kraml:2025fpv}, we outlined the complete one-loop matching of the general Minimal Supersymmetric Standard Model (MSSM) onto the Standard Model EFT (SMEFT) Warsaw basis~\cite{Grzadkowski:2010es}. The MSSM-to-SMEFT matching presents several challenges, the large number of heavy fields being only one of them. One must also account for the subtleties associated with the proper identification of the second Higgs doublet, change in the regularization scheme between the supersymmetry-preserving $\overline{\text{DR}}$ to the SMEFT approach of $\overline{\text{MS}}$, and the non-trivial index contractions due to flavorful sfermion masses. Also, due to R-parity conservation, the tree-level matching conditions only involve the mass of the second Higgs doublet. So, to fully capture the effects of the diverse particle spectrum of the MSSM on low-energy phenomena, it is necessary to conduct the matching at least up to one-loop order. This is why previous results in the literature all applied approximations in order to obtain the MSSM matching conditions (see \cite{Kraml:2025fpv} for references).
In our work, which takes into account all subleading one-loop contributions by all superpartners and including all correlations, we benefitted from automated calculations with the \matchete tool.

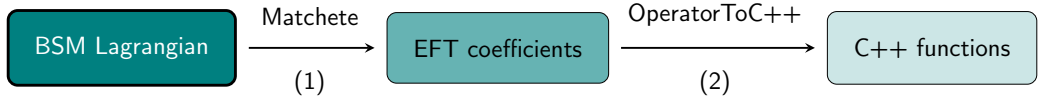
\begin{figure}[t]
\centering
\scalebox{0.9}{\begin{tikzpicture}[node distance=0.3cm and 0.3cm] 

\node (r1) [darkbox, inner sep=11pt] {BSM Lagrangian};
\node (r2) [lightbox, inner sep=11pt, right=2.2cm of r1] {EFT coefficients};
\node (r3) [standardbox, inner sep=11pt, right=3.2cm of r2] {C++ functions};

\draw [arrow, shorten <=5pt, shorten >=5pt] (r1.east) -- (r2.west);
\draw [arrow, shorten <=5pt, shorten >=5pt] (r2.east) -- (r3.west);

\node [font=\sffamily\small, above=0.2cm of $(r1)!0.5!(r2)$] {Matchete};
\node [font=\sffamily\small, below=0.2cm of $(r1)!0.5!(r2)$] {(1)};
\node [font=\sffamily\small, above=0.2cm of $(r2)!0.5!(r3)$] {OperatorToC++};
\node [font=\sffamily\small, below=0.2cm of $(r2)!0.5!(r3)$] {(2)};

\end{tikzpicture}}
\caption{Schematic representation of the steps performed by \matchete and \OpToCpp.}
\label{fig:flowchart-0}
\end{figure}

The matching operation conducted by \matchete is an analytical calculation, represented by step~(1) in Figure~\ref{fig:flowchart-0}. The output involves function wrappers based on \matchete's API. To be able to extract numerical output from these results, it is necessary to unwrap the expressions and recast them as simple function calls. There are additional complexities on account of the fact that there can be (i)~100s\,--\,1000s of individual terms in the matched expressions of some Wilson coefficients, (ii)~a wide-variety of loop-functions encapsulating integrals involving the BSM masses and (iii) variable patterns of repeated index contractions among parameters carrying flavor indices. So, a simple call to Mathematica's inbuilt \texttt{CForm} function is not sufficient to translate the \matchete output to the desired format. This motivated us to adopt an approach that not only ensures fast execution but is amenable to generalisation if higher mass dimensions, higher loop orders, or even non-SMEFT EFTs and non-Warsaw operator bases were to be incorporated in the matching. This led to the development of \OpToCpp, i.e.\ the subject of this paper and step~(2) in Figure~\ref{fig:flowchart-0}. 

\OpToCpp is a hybrid package with both Mathematica and C++ components that automates the conversion of \matchete output into C++ functions. It leverages the object-oriented programming features of C++ to associate the BSM parameters and the Wilson coefficient functions to a class describing the high energy model. It also uses Modern C++ features to implement necessary optimisations and deal with the intricacies associated with the loop-functions and flavor index summations in a generic and extensible way. Additionally, there is also a Python-based frontend which can be used to connect with the wider ecosystem of Python libraries (e.g. \texttt{Wilson} \cite{Aebischer:2018bkb}, \texttt{SMEFiT} \cite{Giani:2023gfq}) and data formats (such as \texttt{WCxf} \cite{Aebischer:2017ugx}) used in EFT based analyses. \OpToCpp is one of the few tools that form a bridge between the symbolic matching and numerical execution in a fast and seamless manner, therefore enabling streamlined phenomenological studies of concrete BSM theories using EFT methods. 
Also, as far as we know, this is the only tool, so far, that translates \matchete output to C++ code. The only numerical export for Wilson coefficient values available in \texttt{Matchete} as of now\footnote{A~Python based interface between \texttt{Matchete} and \texttt{smelli}/\texttt{jelli}~\cite{Aebischer:2018iyb,Smolkovic:2026cba} is currently under development.} 
is given by the \texttt{Matchete} routine \texttt{ExportWCxf}, introduced with~\cite{Belfatto:2025ids}, that allows to directly generate \texttt{WCxf} files for a given BSM Lagrangian. 
For \texttt{MatchmakerEFT}, on the other hand, there exist tools like \texttt{match2fit}~\cite{terHoeve:2023pvs}, which uses the matched expressions to create run cards for \texttt{SMEFiT}, and \texttt{MatchmakerParser}~\cite{Gargalionis:2024jaw}, which converts the results to Python code.

The structure of this article is as follows: we start by giving installation instructions for the package in Section~\ref{sec:install}. Section~\ref{sec:demo} then demonstrates the full workflow for an example model, highlighting the various ways a user can interact with the code. This is followed by a discussion of the notable technical features of \OpToCpp and specific design choices in Section~\ref{sec:technicalities}. Section~\ref{sec:benchmarking} gives some performance benchmarks, and Section~\ref{sec:conclusion} provides a summary and conclusions. The Mathematica and Python functions available to the user are documented in Appendix~\ref{app:doc}, the C++ user interface is described in Appendix~\ref{app:code-listings}.

\section{Getting started}\label{sec:install}

\subsection{Pre-requisites}

\OpToCpp requires the following packages and libraries to be pre-installed locally: 
\begin{enumerate}
    \item \matchete (\texttt{>= v0.3.0}): This is primarily needed to obtain the analytical formulae corresponding to the loop functions in the matching results.

    \item A C++ compiler that abides by the C++17 standard or newer, e.g. \texttt{GCC >= v7.1} and \texttt{Clang >= v4}. For cases involving minimal extensions of the SM, the choice of compiler does not make much difference but for complex scenarios such as the MSSM-to-SMEFT matching, we observed significantly faster compile times with \texttt{Clang} and recommend using it in such cases.

    \item \texttt{meson} + \texttt{ninja}: Our build system comprises of a \texttt{meson.build} file which takes care of cross-platform compilation. The versions used for the most recent builds were \texttt{meson v1.11.1} and \texttt{ninja v1.13.1}.

    \item \texttt{OpenMP}: \texttt{libomp} needs to be locally installed to ensure that repetitive operations such as Einstein Summations across a large number of repeated indices benefit from parallel code execution. If one uses \texttt{Clang} as the compiler (default on MacOS) one would need to install \texttt{libomp}, while \texttt{GCC} comes with \texttt{libgomp} bundled with it. 

    \item \texttt{pybind11} (\texttt{>= v3.0.0}): It provides the necessary headers for building a Python module using the C++ class and functions. 

    \item Python libraries: \texttt{wcxf}, \texttt{pyyaml}, \texttt{numpy}, and \texttt{pandas} are required for the proper functioning of the code provided through the \texttt{utils} module. These can be installed within a Python environment using the \texttt{pip} package manager.
\end{enumerate}
For missing dependencies, \texttt{install.sh} (see below) prints the necessary installation instructions, which, along with other technical details, can also be found in the project \texttt{README}.

\begin{figure}[!htb]
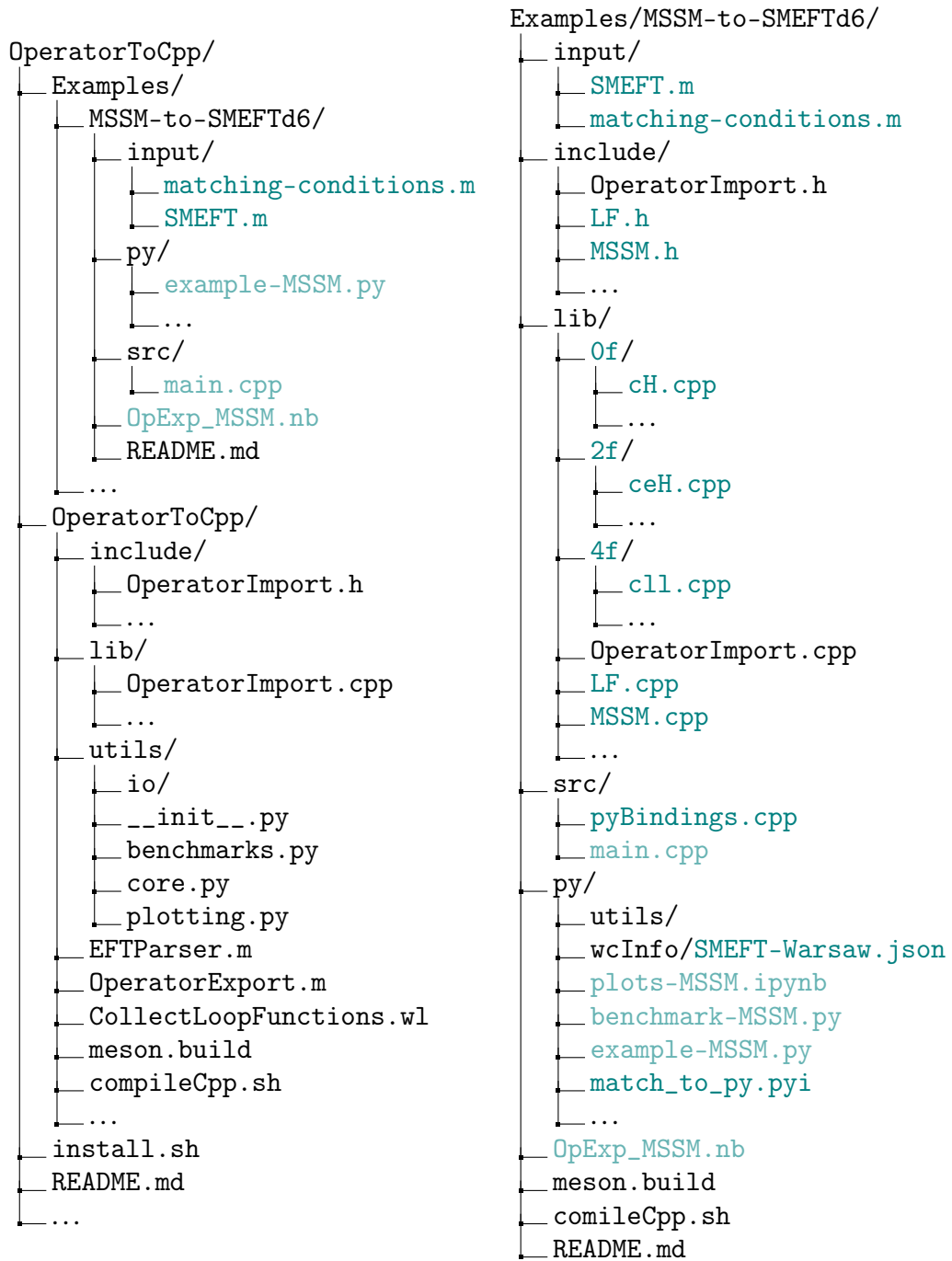

\centering
\begin{minipage}{0.4\textwidth}
    \dirtree{%
    .1 OperatorToCpp/.
    .2 Examples/.
    .3 MSSM-to-SMEFTd6/.
    .4 input/.
    .5 \textcolor{tealL3}{matching-conditions.m}.
    .5 \textcolor{tealL3}{SMEFT.m}.
    .4 py/.
    .5 \textcolor{tealL2}{example-MSSM.py}.
    .5 $\dots$.
    .4 src/.
    .5 \textcolor{tealL2}{main.cpp}.
    .4 \textcolor{tealL2}{OpExp\_MSSM.nb}.
    .4 README.md.
    .3 $\dots$.
    .2 OperatorToCpp/.
    .3 include/.
    .4 OperatorImport.h.
    .4 $\dots$.
    .3 lib/.
    .4 OperatorImport.cpp.
    .4 $\dots$.
    .3 utils/.
    .4 io/.
    .4 \_\_init\_\_.py.
    .4 benchmarks.py.
    .4 core.py.
    .4 plotting.py.
    .3 EFTParser.m.
    .3 OperatorExport.m.
    .3 CollectLoopFunctions.wl.
    .3 meson.build.
    .3 compileCpp.sh.
    .3 $\dots$.
    .2 install.sh.
    .2 README.md.
    .2 $\dots$.
    }
\end{minipage}
\hspace{1cm}
\begin{minipage}{0.4\textwidth}
    \dirtree{%
    .1 Examples/MSSM-to-SMEFTd6/.
    .2 input/.
    .3 \textcolor{tealL3}{SMEFT.m}.
    .3 \textcolor{tealL3}{matching-conditions.m}.
    .2 include/.
    .3 OperatorImport.h.
    .3 \textcolor{tealL3}{LF.h}.
    .3 \textcolor{tealL3}{MSSM.h}.
    .3 $\dots$.
    .2 lib/.
    .3 \textcolor{tealL3}{0f}/.
    .4 \textcolor{tealL3}{cH.cpp}.
    .4 $\dots$.
    .3 \textcolor{tealL3}{2f}/.
    .4 \textcolor{tealL3}{ceH.cpp}.
    .4 $\dots$.
    .3 \textcolor{tealL3}{4f}/.
    .4 \textcolor{tealL3}{cll.cpp}.
    .4 $\dots$.
    .3 OperatorImport.cpp.
    .3 \textcolor{tealL3}{LF.cpp}.
    .3 \textcolor{tealL3}{MSSM.cpp}.
    .3 $\dots$.
    .2 src/.
    .3 \textcolor{tealL3}{pyBindings.cpp}.
    .3 \textcolor{tealL2}{main.cpp}.
    .2 py/.
    .3 utils/.
    .3 wcInfo/\textcolor{tealL3}{SMEFT-Warsaw.json}.
    .3 \textcolor{tealL2}{plots-MSSM.ipynb}.
    .3 \textcolor{tealL2}{benchmark-MSSM.py}.
    .3 \textcolor{tealL2}{example-MSSM.py}.
    .3 \textcolor{tealL3}{match$\_$to$\_$py.pyi}.
    .3 $\dots$.
    .2 \textcolor{tealL2}{OpExp$\_$MSSM.nb}.
    .2 meson.build.
    .2 comileCpp.sh.
    .2 README.md.
    }
\end{minipage}
    \caption{\textit{(left)} Directory structure of the \OpToCpp repository  and \textit{(right)} Expanded working directory for the MSSM-to-SMEFT matching scenario showing (i) model-dependent files that either serve as the input or are generated by \texttt{OperatorExport.m} (in \smallsquare{tealL3}) and (ii) the user interface files (in \smallsquare{tealL2}). Helper files and build artifacts are omitted for brevity.}
    \label{fig:repo-directory}
\end{figure}

\subsection{Download and installation}

\OpToCpp is a free and open-source project. The latest release version can be downloaded from Zenodo \cite{operator-to-cpp:github}.\footnote{The entire code, development history, and open issues are publicly available on the GitHub repository \url{https://github.com/BSM-EFT/OperatorToCpp}.} After extracting the downloaded \texttt{(.zip)} file, one gets access to the directory structure shown on the left side of Figure~\ref{fig:repo-directory}. The core components of the package are stored in the \texttt{OperatorToCpp/} directory. These consist of
\begin{enumerate}
    \item \texttt{OperatorExport.m} -- A Mathematica package that provides functions to systematically translate \matchete output to C++ classes and methods and generates model-specific \texttt{.h} and \texttt{.cpp} files.

    \item \texttt{EFTParser.m} -- A Mathematica package that provides simple functions to read EFT metadata and store them in JSON format.

    \item \texttt{CollectLoopFunctions.wl} -- A Mathematica script that extracts all unique loop functions (defined in \matchete's convention), their analytical forms, and degenerate limits and stores them into source (\texttt{LF.cpp}) and header (\texttt{LF.h}) C++ files.

    \item \texttt{OperatorImport.cpp} -- A C++ source file that defines helper functions to enable computations such as loop-function evaluation, Einstein-summation, etc., within the exported expressions. The corresponding header file is stored in the include directory.

    \item \texttt{utils} -- A Python module designed to provide file input-output facilities and to interface with popular formats such as \texttt{csv}, \texttt{yaml}, \texttt{WCxf} etc.
\end{enumerate}
It also contains template \texttt{meson.build} and \texttt{compileCpp.sh}, i.e.\ model-independent build-system files that enable the compilation of the code into executables and a Python module.

The \texttt{Example/} directory contains template working directories for a few scenarios, and provides input files with the matching results and EFT information, along with sample user-interface files in the form of (i) a Mathematica notebook ( \texttt{OpExp\_<model>.nb}) (ii) a C++ source file \texttt{main.cpp} and (iii) Python scripts \texttt{example-<model>.py}. 

The relevant (package and build system) files can be installed into the working directory using the \texttt{install.sh} script. For the MSSM-to-SMEFT matching results\cite{Kraml:2025fpv}, one must execute the following commands in a terminal:
\begin{quote}
    \texttt{./install.sh Examples/MSSM-to-SMEFTd6}\\
    \texttt{cd Examples/MSSM-to-SMEFTd6} 
\end{quote}
The C++ files can then be generated by evaluating the commands in the Mathematica notebook OpExp\_MSSM.nb. Finally, by executing the compileCpp.sh script from the working directory, one can build the \texttt{match\_to\_py} Python module (and any C++ executables). This module contains the MSSM model class with the model parameters as member variables and the SMEFT Wilson coefficients as its methods. 

The directory structure after all C++ and Python files have been generated is shown on the right side of Figure~\ref{fig:repo-directory}; the complete workflow for transpiling the MSSM-to-SMEFT matching results will be discussed in detail in Section~\ref{sec:demo}.

\section{Interacting with the code}\label{sec:demo}

\begin{figure}[!htb]
\hspace{1.5cm}\scalebox{0.73}{
\begin{tikzpicture}[node distance=0.3cm and 0.3cm] 
\node (r1) [lightbox, inner sep=9pt] {EFTParser.m};
\node (r2) [lightbox, inner sep=9pt, below=of r1] {CollectLoopFunctions.wl};
\node (r3) [lightbox, inner sep=9pt, below=of r2] {OperatorExport.m};
\node (groupbox1) [group_dash, fit=(r1)(r2)(r3)] {};
\node (r4) [standardbox, inner sep=9pt, left=2cm of r2]{OpExp$\_$MSSM.nb};
\node (r5) [darkbox, inner sep=9pt, left=2cm of r4, yshift=0.65cm]{SMEFT.m};
\node (r6) [darkbox, inner sep=9pt, below=of r5]{matching-conditions.m};
\node (groupbox2) [group_dash, fit=(r5)(r6)] {};
\node (r8) [darkbox, inner sep=9pt, below=3.5cm of r4]{MSSM.h(.cpp)};
\node (r10) [darkbox, inner sep=9pt, left=of r8]{LF.h(.cpp)};
\node (r9) [darkbox, inner sep=9pt, right=of r8]{cXX.cpp};
\node (r11) [darkbox, inner sep=9pt, below=of r8]{pyBindings.cpp};
\node (r12) [lightbox, inner sep=9pt, below=of r11]{OperatorImport.h(.cpp)};
\node (groupbox3) [group_dash, fit=(r8)(r9)(r10)(r11)(r12)] {};
\node (r13) [lightbox, inner sep=9pt, below=9.5cm of r4]{meson.build};
\node (r14) [darkbox, inner sep=9pt, below=2cm of r13]{match$\_$to$\_$py$<\cdots>$.so};
\node (r15) [darkbox, inner sep=9pt, left=of r14]{SMEFT-Warsaw.json};
\node (r16) [lightbox, inner sep=9pt, right=of r14]{utils};
\node (groupbox4) [group_dash, fit=(r14)(r15)(r16)] {};
\node (r17) [standardbox, inner sep=9pt, left=3cm of r13]{main.cpp};
\node (r19) [standardbox, inner sep=9pt, below=2cm of r14]{example-MSSM.py};
\node (r18) [darkbox, inner sep=9pt, right=3.6cm of r19] {C++ executable};
\node (r20) [font=\sffamily\small, below=2cm of r19, text width=6cm,align=center] {Numerical calculations in Python; interface with wcxf, yaml, csv formats};
\node (r21) [font=\sffamily\small, below=2cm of r18, text width=6cm,align=center] {Numerical evaluation in C++};
\draw [arrow] (groupbox1.west) -- (r4.east);
\draw [arrow] (groupbox2.east) -- (r4.west);
\draw [arrow] (r4.south) -- (groupbox3.north);
\draw [arrow] (groupbox3.south) -- (r13.north);
\draw [arrow] (r13.south) -- (r14.north);
\draw [arrow] (r17.east) -- (r13.west);
\draw [arrow] (r13.east) -- ++(5.4,0)  -- (r18.north);
\draw [arrow] (r14.south)+(0,-0.3) -- (r19.north);
\draw [arrow] (r19.south) -- (r20.north);
\draw [arrow] (r18.south) -- (r21.north);
\end{tikzpicture}}
\caption{A flowchart describing the complete \OpToCpp workflow, that converts the results of MSSM-to-SMEFT matching into a C++ class and its methods, which can be used to build a C++ executable or a Python library. The differently coloured blocks represent (i) the user-interface in \smallsquare{tealL1}, (ii) the core components of \OpToCpp in \smallsquare{tealL2}, and (iii) the input and the model-dependent files generated at different steps in \smallsquare{tealL3}. 
} 
\label{fig:flowchart}
\end{figure}
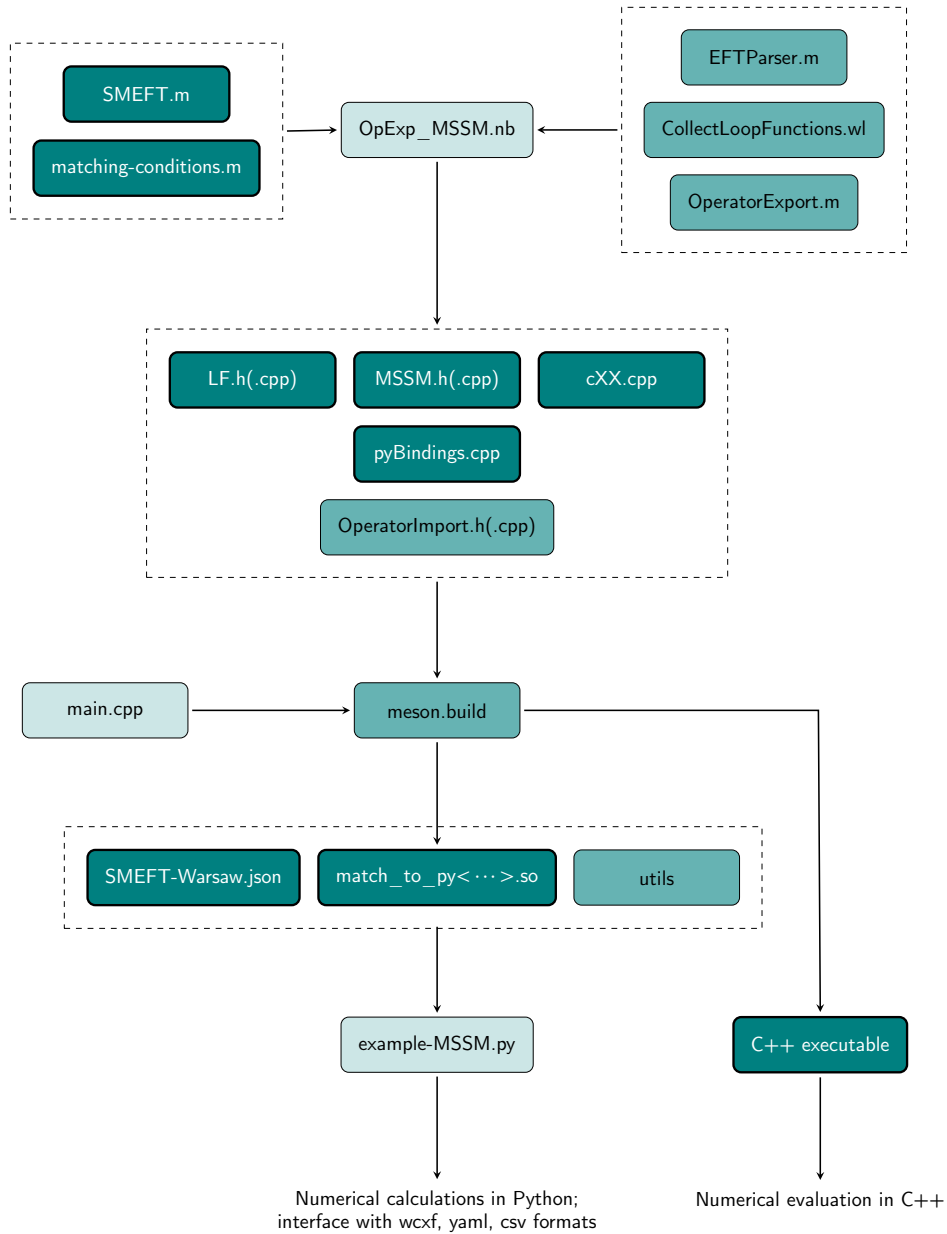

We will now describe the different stages where the user can interact with the code using the MSSM-to-SMEFT matching as an explicit example. Figure~\ref{fig:flowchart} schematically illustrates how the core \OpToCpp files process the input (\matchete results and EFT information), to generate model-dependent files at three distinct steps, through a Mathematica notebook, a C++ source file or a Python script.

\subsection{Mathematica user interface}

The Mathematica component of \OpToCpp admits \textbf{input} in the form of
\begin{enumerate}
    \item[(i)] The EFT model file used to define the operator basis and  coefficients for \matchete. For the current case, \texttt{"SMEFT.m"} describes the dimension-6 SMEFT coefficients in the Warsaw basis.
    
    \item[(ii)] A file storing the MSSM-to-SMEFT matching relations as a list of rules with the EFT coefficients as keys and the matched expressions as values. This is shown in Figure \ref{fig:flowchart} as \texttt{"matching-conditions.m"}.
\end{enumerate}
\noindent The processing of the input involves:
\begin{enumerate}
    \item \underline{Parsing EFT information}: The \texttt{EFTParser.m} package provides functions to read the Wilson coefficient information from \texttt{SMEFT.m} and store them in a JSON file for further use. This package relies on the \matchete function \texttt{GetCouplings}.

    \item \underline{Extracting loop-functions}: The script \texttt{CollectLoopFunctions.wl} reads the contents of \texttt{matching-conditions.m}, extracts the unique loop functions and their degeneracy limits, 
    and translates all expressions to C++ friendly form. The \matchete function \texttt{EvaluateLoopFunctions} is used to obtain the analytical forms of loop functions.

    \item \underline{Converting matching relations to C++ expressions}: The final and main part of the workflow involves a systematic translation of the matching results and their export to C++. This is enabled by the \texttt{OperatorExport.m} package.  
\end{enumerate}

\begin{figure}[!htb]
\begin{lstlisting}[style=mathematicaMargin,caption={A snippet of the Mathematica user interface, \texttt{OpExp\_MSSM.nb}, for translating the matching results for the MSSM.}, label={lst:OpExp.nb}]

?\hspace{-0.6cm}\textcolor{black!40!red}{\small Read EFT Metadata}?

?\InPromptNext? SetDirectory[NotebookDirectory[]];
?\InPromptNext? AppendTo[$Path, FileNameJoin[
    {ParentDirectory[NotebookDirectory[], 2], "OperatorToCpp"}]];

?\InPromptNext? Needs["EFTParser'"]
?\InPromptNext? WCInfo = ReadEFTInfo["SM", "./input/SMEFT.m"];
?\InPromptNext? ExportWCInfo["SMEFT", "Warsaw", WCInfo]

?\hspace{-0.6cm}\textcolor{black!40!red}{\small Load the saved output and extract loop-functions}?

?\InPromptNext? $MatchedResultsPath = FileNameJoin[{NotebookDirectory[], 
        "input", "matching-conditions.m"}];
 Get["CollectLoopFunctions.wl"];
?\InPromptNext? matchedResult = Import[$MatchedResultsPath];

?\hspace{-0.6cm}\textcolor{black!40!red}{\small Load the OperatorExport module}?

?\InPromptNext? Needs["OperatorExport'"]
?\OutPromptOfCurrent? ?\hspace{0.05cm}\textcolor{black!70!gray}{\sffamily OperatorExport v1.0.0 - by Suraj Prakash (\textcolor{blue}{suraj.prakash@ific.uv.es})}?
?\hspace{0.25cm}\textcolor{black!70!gray}{\sffamily Affiliation: IFIC (Universitat de Valencia - CSIC)}?
?\hspace{0.25cm}\textcolor{black!70!gray}{\sffamily Github: \textcolor{blue}{https://github.com/BSM-EFT/OperatorToCpp}}?
?\hspace{0.25cm}\textcolor{black!5!gray}{\sffamily  Mathematica component of the OperatorToC++ code.}?

?\hspace{-0.6cm}\textcolor{black!40!red}{\small Rename variables (model-dependent step) and simplify}?

?\InPromptNext? ReplaceTrigCouplings = {?$s\gamma \rightarrow (1 - c\gamma^2)^{1/2}$?, ?$c2\gamma \rightarrow (2 c\gamma^2 - 1)$?, ?$s2\gamma \rightarrow 2(1 - c\gamma^2)^{1/2}c\gamma$?, 
   ... };
?\InPromptNext? variableReplacement = { 
   ?$\mu$?bar2 ?$\rightarrow$? mubarsq, m?$\Phi\rightarrow$? mPhi, ?$\mu$?t ?$\rightarrow$? muTilde, c?$\gamma$? ?$\rightarrow$? cgamma,
   cHqu ?$\rightarrow$? Yu, cHqd ?$\rightarrow$? Yd, cHle ?$\rightarrow$? Ye, cH2 ?$\rightarrow$? (-mHsq), 
   cB2 ?$\rightarrow$? g1, cW2 ?$\rightarrow$? g2, cG2 ?$\rightarrow$? g3};

?\InPromptNext? matchedResult = matchedResult /. ReplaceTrigCouplings /. variableReplacement
?\InPromptNext? SimplifiedOutput = SimplifyOutput[matchedResult, WCInfo];

?\hspace{-0.6cm}\textcolor{black!40!red}{\small Parameter extraction and file creation}?

?\InPromptNext? ComplexParams = {ad, ae, au, Yd, Ye, Yu};

?\InPromptNext? BuildFiles["MSSM", SimplifiedOutput, ComplexParams, WCInfo]
?\hspace{0.25cm}\textcolor{black!30!gray}{\sffamily C++ header (.h) files successfully created and placed in the ./include/ subdirectory.}?
?\hspace{0.25cm}\textcolor{black!30!gray}{\sffamily C++ source (.cpp) files successfully created and placed in the ./lib/ subdirectory.}?
?\hspace{0.25cm}\textcolor{black!30!gray}{\sffamily pyBindings.cpp successfully created and placed in the ./src/ subdirectory.}?
?\hspace{0.25cm}\textcolor{black!30!gray}{\sffamily match\_to\_py.pyi successfully created and placed in the ./py/ subdirectory.}?
\end{lstlisting}
\end{figure}
\noindent These steps are executed through the \texttt{OpExp\_MSSM.nb} notebook, whose contents are reproduced in Listing~\ref{lst:OpExp.nb}.
Some explanations are in order:
\begin{enumerate}
    \item First, we add \OpToCpp to path, so that the included Mathematica packages and scripts can be imported conveniently.
    
    \item The second step involves loading the \texttt{EFTParser} package, reading the \texttt{SMEFT.m} file using the \texttt{ReadEFTInfo}, and storing the Wilson coefficient attributes, such as the number of fermion flavours, self-conjugate nature, and their symmetries in a dictionary (\texttt{`WCInfo'} in this example). The \texttt{ExportWCInfo} function places the contents of this dictionary in the \texttt{py/wcInfo/SMEFT-Warsaw.json} file.
    
    \item Then, \texttt{CollectLoopFunctions.wl} is executed to extracts unique loop functions, and their degeneracy limits. It then creates \texttt{include/LF.h} and \texttt{lib/LF.cpp} to declare and define a C++ wrapper for loop functions, while individual expressions are exported to \texttt{LF\_helper.cpp(.h)} as separate functions.

    \item Next, we load the \texttt{OperatorExport} package which provides the necessary functions to simplify and export the matching results to C++ files.

    \item The \texttt{SimplifyOutput} command converts the matching results to a simplified intermediate form, stripped of the \matchete API (but not yet ready for export to C++). However, before this step, we recommend applying model-dependent symbol and(or) variable replacements. 
    
    \textsf{Cell~[9]}
    in Listing~\ref{lst:OpExp.nb} shows a replacement rule based on relations among trigonometric couplings, which are obfuscated in the matching results. Similarly, the first row in 
    \textsf{Cell~[10]}
    shows a renaming of variable names containing Greek letters to enable a smooth translation to C++ code, while the last two rows in \textsf{Cell~[10]} show a renaming of the Yukawa, squared Higgs mass and gauge parameter in the SMEFT basis.

    \item In  
    \textsf{Cell~[13]}, 
    we explicitly specify which parameters are complex. This is needed to prevent double counting of such variables and their conjugates, while also ensuring that their conjugates are properly constructed. 

    \item Finally, the \texttt{BuildFiles} function creates the following files:
        \begin{enumerate}
            \item[(i)] \texttt{include/MSSM.h} and \texttt{lib/MSSM.cpp}, i.e., C++ header and source files, that declare and define the \texttt{MSSM} class, constructor, updater, getter and setter methods. The Wilson coefficient methods are declared in \texttt{MSSM.h} but their definitions are stored in separate \texttt{cXX.cpp} placed within subdirectories \texttt{lib/0f/}, \texttt{lib/2f/}, \texttt{lib/4f/} based on the number of fermions in the corresponding operator;\footnote{This is done to optimise the compilation process.}

            \item[(ii)] the \texttt{src/pyBindings.cpp} file that enables the use of the \texttt{MSSM} class and the Wilson coefficient methods from a Python frontend, using the \texttt{pybind11} library;

            \item[(iii)] the \texttt{py/match\_to\_py.pyi} file which stores the declarations of a Python class named \texttt{MSSM} along with its Wilson coefficient methods. 
        \end{enumerate}
    A snippet of the contents of \texttt{MSSM.h} is shown in Appendix~\ref{app:code-listings}.
\end{enumerate}

\subsection{C++ backend and Python frontend}

After having generated the C++ header and source files from the Mathematica code, one can proceed further in one of two ways:
\begin{enumerate}
    \item The model (e.g. MSSM) class and the Wilson coefficient methods can be accessed from a C++ source file, shown as \texttt{main.cpp} in Figure~\ref{fig:flowchart}. User interaction with the \texttt{MSSM} class and its methods through a C++ source file is described in more detail in Appendix~\ref{app:code-listings}. Having access to the Wilson coefficients as C++ functions makes it possible to connect with tools and libraries such as \texttt{HEPfit} \cite{DeBlas:2019ehy} for numerical analyses. 
    
    However, for most users we recommend the second approach, of interacting with the compiled functions from a Python-based front-end, described next. 

    \item By compiling the C++ files generated by \texttt{OpExp\_MSSM.nb}, e.g. \texttt{MSSM.cpp}, \texttt{LF.cpp}, \texttt{pyBindings.cpp} etc.,  along with \texttt{OperatorImport.cpp}, one can create a shared object (\texttt{.so}) file which furnishes the \texttt{match\_to\_py} Python module. This module and the \texttt{match\_to\_py.pyi} file enable the user to import the \texttt{MSSM} class and call the Wilson coefficient methods from within a Python program, see Listing~\ref{lst:example.py}.

    In this approach, one retains the speed and efficiency of the compiled C++ code while also enjoying the ease of use and ease of development offered by Python.
\end{enumerate}

\begin{figure}[!htb]
\begin{lstlisting}[style=Python, caption={A simple Python script that (i) creates an instance of the MSSM model and evaluates Wilson coefficients, (ii) demonstrates some file IO facilities provided by the \texttt{utils} module, and (iii) generates a simple lineplot of Wilson coefficients for a varying model parameter.}, label={lst:example.py}]
    from match_to_py import MSSM
    from utils.io import write_to_wcxf
    from utils.core import eval_wc
    from matplotlib import pyplot as plt
    import numpy as np
    
    # define a parameter-dictonary
    param_dict = { "g1": 0.37, "g3": 1.1, "cgamma": 0.01, "m1":1.2e3, ...,
                   "mut1":1e9, "mut2":1e9, "mut3":2e3, ...,
                   "Yu11":1e-5, ..., "Yu33":0.9 }

    # initialize an instance of the MSSM model with the parameter dictionary,
    # renormalization scale set to 1000 GeV and loop contrbutions turned on
    model1 = MSSM(param_dict, 1e3, True)

    # evaluate Wilson coefficients as a method call
    print(model1.cG())
    print(model1.cuB(2, 2))

    
    # evaluate coefficients and write to a WCxf file

    wcs = ["uu_1331", "G", "phiG", "phiD", "uG_33"]
    eft_info = { "eft": "SMEFT", "basis": "Warsaw" }
    write_to_wcxf("example_wcxf.yaml",model1,eft_info,wcs,opt="seq")
    write_to_wcxf("example_wcxf_all.yaml",model1,eft_info)

    # obtain arrays of Wilson coefficients for varying m1, 
    # keeping other parameters fixed

    m1_range = np.linspace(1200,2700,16)
    cqq1_vals, cuG_vals, cHq1_vals = [], [], []
    for m in m1_range:
        d = {"m1": m}
        model1.updateParams(d)
        cqq1_vals.append(eval_wc(model1, "cqq1_3333"))
        cuG_vals.append(eval_wc(model1, "cuG_33"))
        cHq1_vals.append(eval_wc(model1, "cHq1_33"))

    # Set plot attributes
    plt.rcParams['axes.labelsize'] = 20
    plt.rcParams["text.usetex"] = True
    ...
    ...

    # Create the plot 
    plt.figure(figsize=(8,8))
    plt.plot(m1_range/1e3, np.abs(cqq1_vals)*1e12, color="red",
             label=r"$|C_{qq}^{(1),3333}|$")
    plt.plot(m1_range/1e3, np.abs(cuG_vals)*1e12, color="blue",
             label=r"$|C_{uG}^{33}|$")
    plt.plot(m1_range/1e3, np.abs(cHq1_vals)*1e12, color="green",
             label=r"$|C_{Hq}^{(1),33}|$")
    plt.xticks([1.2,1.5,1.8,2.1,2.4,2.7])
    plt.yticks([1,2,3,4,5,6])
    plt.xlim(1.2,2.7)
    plt.ylim(1,6)
    plt.xlabel(r"$m_1\,\,[{\rm TeV}]$")
    plt.ylabel(r"$10^{-12}\,\,[{\rm GeV}^{-2}]$")
    plt.legend()
    plt.savefig("lineplot.pdf")
\end{lstlisting}
\end{figure}

\clearpage

\paragraph{The build process:} To aid in compilation, we provide a template \texttt{meson.build} file. It specifies two build targets, a C++ executable and the Python module, 
\begin{quote}
    \texttt{executable(`main.out', `src/main.cpp', \dots)}\\
    \texttt{py.extension\_module(`match\_to\_py', `src/pyBindings.cpp', \dots)}
\end{quote}
If the source file is \textbf{not} named \texttt{main.cpp} or if there is more than one source file, the specification of the C++ executable needs to be modified accordingly. In case there is no C++ executable to be built, the corresponding command should be removed from \texttt{meson.build}. 
The commands for building and installing the targets are collected in the script \texttt{compileCpp.sh}. Executing the command \hspace{0.2cm}\texttt{"./compileCpp.sh"}\hspace{0.2cm} in a terminal from the working directory (\texttt{Examples/MSSM-to-SMEFTd6} in this case) compiles all source files, links them and the dependencies (\texttt{libomp}, \texttt{pybind11}) and places the build artifacts in the install directories specified in \texttt{meson.build}. By default, the C++ executables are placed in the working directory, and \texttt{.so} file in the \texttt{py/} subdirectory.\footnote{If the directory structure of the C++ files is modified, then parts of the \texttt{meson.build} file will require modification accordingly. Therefore, maintaining the directory structure shown in Figure~\ref{fig:repo-directory} is strongly recommended.}

\paragraph{The user interface:} With the \texttt{match\_to\_py} shared object (\texttt{.so}) and Python interface (\texttt{.pyi}) files at disposal, one can use them like any other Python library and import the model and call its Wilson coefficient methods from a Python (\texttt{.py}) file or a Jupyter (\texttt{.ipynb}) notebook. Listing~\ref{lst:example.py} shows how to import the \texttt{MSSM} class from the \texttt{match\_to\_py} module, initialize an instance of this class, invoke its Wilson coefficient methods, interact with the functions defined in the \texttt{utils} module, and produce a simple plot.

\begin{figure}[t]
    \centering
    \includegraphics[scale=0.5]{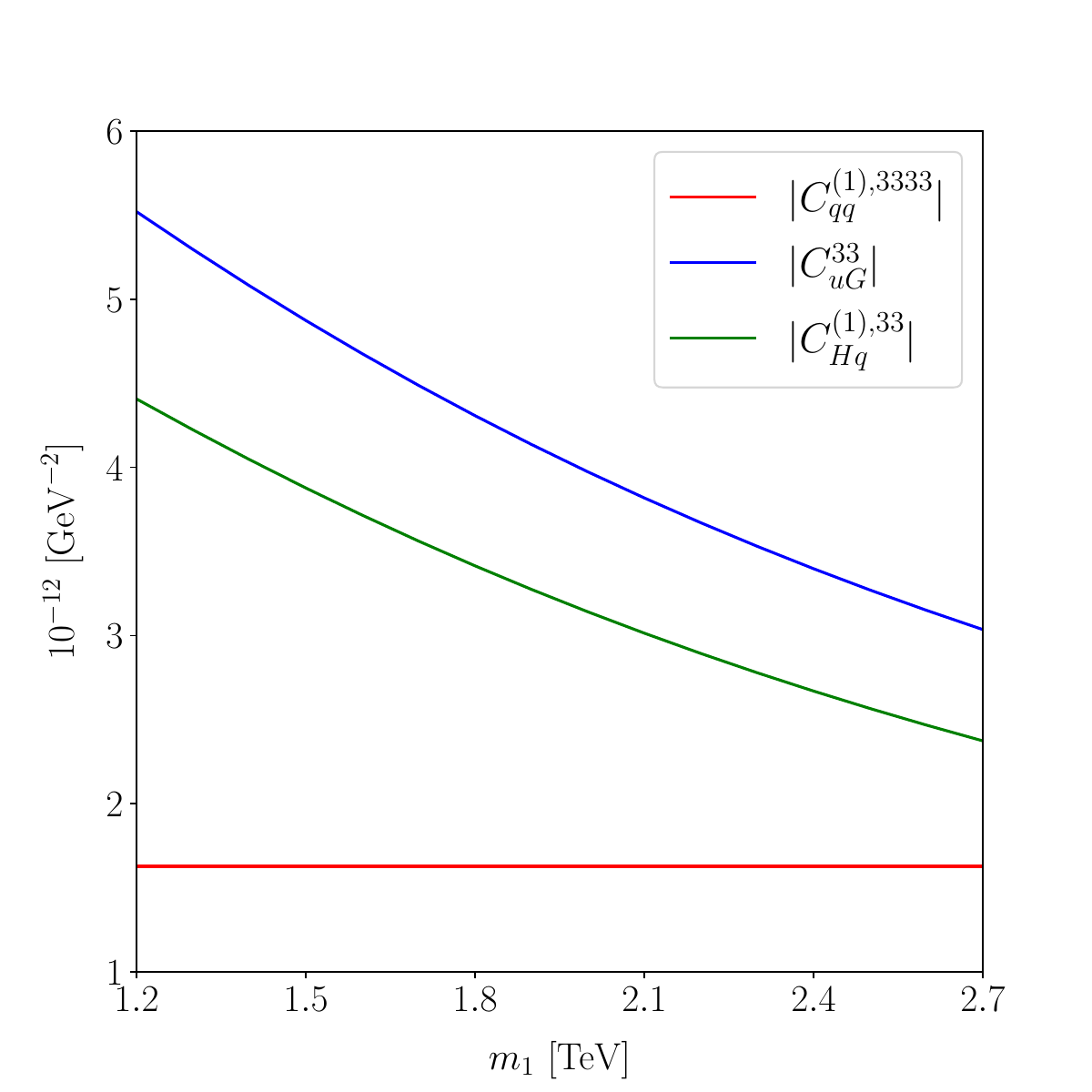}
    \caption{Line-plot generated by the plotting commands in Listing~\ref{lst:example.py}, showing the variation of the Wilson coefficients $C_{qq}^{(1),3333}$, $C_{uG}^{33}$, $C_{Hq}^{(1),33}$ with respect to $m_1$, based on the MSSM-to-SMEFT matching relations.}
    \label{fig:wc-lineplot}
\end{figure}

The constructor \texttt{MSSM(param\_dict:\;dict, scale:\;float, loop:\;bool)} allows us to initialize a model instance using a parameter dictionary, a renormalization scale and a boolean specifying if loop corrections are to be included or not. The dictionary contains strings representing model parameters as keys, and floating-point or complex numbers as values. The keys should match the symbols used to define the model parameters in the \matchete input or their replacements specified in \texttt{OpExp\_MSSM.nb} (see Listing~\ref{lst:OpExp.nb}, \textsf{Cell[10]}). 
For vector and matrix-valued parameters, each element must be specified separately. Therefore, the keys must contain number(s), specifying the element, suffixed to the parameter name, e.g.,
\begin{quote}
  \{ \texttt{"mut1"}, \texttt{"mut2"}, \texttt{"mut3"} \} \qquad\,\,\text{for}\quad \{ \texttt{mut[0]}, \texttt{mut[1]}, \texttt{mut[2]} \} \quad\text{and}\\
  \{ \texttt{"Yu11"}, \texttt{"Yu12"}, ..., \texttt{"Yu33"} \} \quad\text{for}\quad \{ \texttt{Yu[0][0]}, \texttt{Yu[0][1]}, ..., \texttt{Yu[2][2]} \}
\end{quote}
Complex conjugates are defined internally and unspecified parameters are initialized to zero. 
To check or update parameter values, the renormalization scale and the inclusion (or exclusion) of loop contributions, each model class generated using \OpToCpp contains the following additional methods:
\begin{itemize}
    \item  \texttt{updateParams({\color{teal}param\_dict:}\;{\color{codekw}dict[string, float|complex]})} enables updating the values of one or more parameters by admitting a dictionary with name-value pairs.
    \item \texttt{getParams()} returns a dictionary containing name-value pairs for each parameter based on the current state.
    \item \texttt{getScale()} returns the value of the renormalization scale.
    \item \texttt{setScale({\color{teal}scale:}\;{\color{codekw}float})} sets the value to the specified scale.
    \item \texttt{loopContributions({\color{teal}True|False})} can be used to toggle on and off, the terms arising from loop-level matching.
\end{itemize}
The \texttt{utils.io} submodule provides file read-write features by interfacing with different file formats. For instance, the \texttt{write\_to\_wcxf} function enables the export of all (or a specified set of) evaluated coefficients to a \texttt{.yaml} file in \texttt{WCxf} format. The bottom half of Listing~\ref{lst:example.py} shows how seamlessly the model class and its methods can interact with Python libraries like \texttt{numpy} and \texttt{matplotlib} and generate simple plots, see Figure~\ref{fig:wc-lineplot}. The release package contains an additional Jupyter notebook \texttt{plots-MSSM.ipynb}, which demonstrates how the \texttt{MSSM} class can be used to produce the 2d and bar plots from Figures~1--3 in Ref.~\cite{Kraml:2025fpv}.

\paragraph{Providing user input and interpreting the output:}
It is important to understand the exact role of \OpToCpp within a phenomenological analysis. \OpToCpp translates the symbolic Mathematica output generated by \matchete to classes, methods, and expressions that can be conveniently interfaced with C++ and Python code, so that one can provide numerical input for the model parameters and obtain numerical output for the Wilson coefficients. However, it does not do any validation or provide any pre-processing facilities for the parameter input. Therefore, the user must keep in mind the following when preparing the numerical input:
\begin{itemize}
    \item The numerical values of the model parameters must be given in the basis employed by \matchete for the matching, i.e., the heavy-particle mass basis. This is not equal to the SM fermion mass basis in general, that is the SMEFT Wilson coefficients are generally not in the up- or down-aligned basis as, e.g., in the predefined bases of the \texttt{WCxf} format.

    \item There is no internal evaluation of parameters based on an (electroweak) input scheme. It is left to the user to ensure that the low-scale SM parameters are obtained correctly for a given set of high scale inputs.

    \item We do not conduct any renormalization group (RG) evolution of parameters. Hence, the SM parameter values must be provided at the matching scale and not at $m_Z$ or similar low-energy scales.

    \item The notion of GeV or TeV units  is not enforced internally. The only relation that holds is that the Wilson coefficients carry inverse-squared powers of the units of mass. The interpretation of the units of the input and output values is up to the user. 

    Consistency is important here. For instance, in Listing.~\ref{lst:example.py}, the masses \texttt{m1}, \texttt{mut3} and the renormalization scale have the values, 1200, 2000 and 1000, respectively. If one assumes that these are in units of GeV, then the output of the Wilson coefficients will be in units of GeV${}^{-2}$. It is perfectly valid to use the values \texttt{m1:\;1.2},\texttt{ mut3:\;2.0} and\texttt{ scale:\;1.0} and interpret these as values in units of TeV. The Wilson coefficients should then be interpreted as TeV${}^{-2}$. However, in any choice of units, all dimensionful quantities must be scaled consistently. For instance, it would be erroneous to input \texttt{m1:\;1.2},\texttt{ mut3:\;2000}~and interpret the former in TeV and the latter in GeV\,.

    Overall, we recommend working in  (implicit) units of GeV as  these are the units in which the \texttt{WCxf} output is generally presented.
\end{itemize}

Regarding the output values for the Wilson coefficients, we must emphasize that:
\begin{itemize}
    \item The values of the Wilson coefficients are evaluated at the matching scale and should therefore be RG-evolved to the appropriate scale(s) using, for example, codes such as \texttt{wilson}~\cite{Aebischer:2018bkb} or \texttt{DsixTools}~\cite{Fuentes-Martin:2020zaz}, before incorporating them within low-energy observables.
    \item The \texttt{write\_to\_wcxf} function generates output in the \texttt{`Warsaw'} basis for the \texttt{`SMEFT'} EFT defined in the \texttt{WCxf} database. In this basis, the down-type quark mass matrix is diagonal at the matching scale.\footnote{Note that a consistent \texttt{WCxf} output is only generated if the basis used for the matching, i.e., the heavy mass eigenbasis, can be made to coincide with the down-quark mass basis. For many theories this is possible, however, not so for the MSSM where the matching is performed in the sfermion mass basis which is generally different from the down-quark mass basis as long as the sfermion masses are not aligned to the Yukawas (See the discussion at the end of Section~2.2.2 in Ref.~\cite{Kraml:2025fpv} for details). In such cases one must include the appropriate unitary matrices in the matching conditions for the flavorful Wilson coefficients that rotate from the heavy mass eigenbasis to the down-aligned basis, which is left to the user.} Also, it follows the convention where coefficients such as  $C_{ll}$ are defined in a non-redundant manner. As a consequence, we have, e.g., 
    $$\texttt{cll[1,2,2,1]} = \# \quad \textrm{and}\quad \texttt{cll[2,1,1,2]} = 0,$$
    whereas \matchete uses the symmetric basis with
    $$\texttt{cll[1,2,2,1]} = \#/2 \quad \textrm{and}\quad \texttt{cll[2,1,1,2]} = \#/2.$$
    For more details see  Appendix C.3 of~\cite{Fuentes-Martin:2020zaz}.
    \item It is possible to evaluate and write Wilson coefficients to file, entirely in the \matchete convention. For this one can use the native \texttt{yaml} file writer function \texttt{write\_to\_yaml}, also provided in \texttt{utils.io}, see Appendix \ref{app:doc} for details.
\end{itemize}
The \texttt{WCxf} output generated for the model defined in Listing~\ref{lst:example.py} was validated against the similar output generated from  the matching relations using the \matchete function \texttt{ExportWCxf}. Additionally, we also validated the \texttt{WCxf} output for simpler UV models, such as, a Singlet Scalar Extension of the SM~\cite{Haisch:2020ahr} and the Type I Seesaw model~\cite{Zhang:2021jdf}. The Python scripts that generate the \texttt{WCxf} output for these models are available in dedicated directories in the release package, i.e., 
\texttt{Examples/SSE-to-SMEFTd6/py} and \texttt{Examples/SeesawTypeId6/py} respectively. 
%

\section{Technical design}\label{sec:technicalities}
This section sheds light on some relevant aspects of the \OpToCpp technical design. It also highlights how a number of challenges were addressed and why certain choices were made during the development process.

\subsection{Loop functions}

When computing the one-loop matching conditions for the Wilson coefficients of an EFT, in terms of the parameters of a UV model, one inevitably encounters integrations involving loop-momenta. In the output generated by \matchete, such integrals are encapsulated within loop-functions, i.e. \texttt{LF} calls, which are defined as
\begin{eqnarray}
\texttt{LF}[\{m_1, m_2, ...\}, \{a, b, ... , p\}] = \int\frac{d^4 k}{(2\pi)^4} \frac{1}{(k^2 - m_1^2)^a (k^2 - m_2^2)^b \dots k^{2p}}.
\end{eqnarray}
Their analytical expressions can be obtained by calling the \texttt{EvaluateLoopFunctions} function provided by \matchete. However, explicitly evaluating the loop-functions before exporting them to C++ introduces two problems,
\begin{enumerate}
    \item[(i)] The number of terms in the full expanded expression becomes much larger and we lose the \textit{relatively} compact, functional structure of the original output.

    \item[(ii)] The analytical expressions generated by \texttt{EvalutaLoopFunctions} can contain terms that lead to numerical instabilities in case of mass-degeneracies, i.e.\ in the limit $m_i \rightarrow m_j$. 
    Degeneracy in the symbolic expression is handled automatically by \matchete by reduction to a simpler \texttt{LF} call, e.g.,
    \begin{eqnarray}
        \texttt{LF}[\{m_1, m_1, m_2\}, \{1,1,1,0\}] \rightarrow \texttt{LF}[\{m_1, m_2\}, \{2,1,0\}].
    \end{eqnarray}
    However, the real issue arises when the masses of two different particles are set to (nearly) the same values in the user input or in a scan.
\end{enumerate}
To address these issues, we do not  evaluate any loop functions before exporting to C++. Also, as shown at the beginning of Listing~\ref{lst:OpExp.nb}, we extract all unique loop-functions within the matching results, using the \texttt{CollectLoopFunctions.wl} script. For each case, this script collects the evaluated expressions as well as all possible reduced loop functions (taking into account the various degeneracy limits), and stores this information in the files \texttt{LF.cpp} (and \texttt{LF\_helper.cpp}). In the current version of the \OpToCpp, we treat the scenario where two masses are within a small percentage of each other, as an exact numerical degeneracy. This is done to avoid large numbers coming from small mass differences in the denominator, or very small arguments in the logarithms. This means we do approximations  such as
\begin{eqnarray}
        \texttt{LF}[\{m_1, m_2, m_3\}, \{1,1,1,0\}]\bigg|_{1 - \epsilon < m1/m2 < 1 + \epsilon} \simeq \texttt{LF}[\{m_1, m_3\}, \{2,1,0\}].
    \end{eqnarray}
at the C++ end. By default,  $\epsilon$ is set to 0.02 internally, using the `\texttt{\#define TOL 0.02}` macro at the top of \texttt{LF.cpp}.\footnote{In future versions, we intend to make this reduction more robust, e.g.\ by resorting to Taylor expansions, as done in \texttt{Matchete}'s \texttt{ExportWCxf} routine.}

While this extraction of loop-functions and the creation of \texttt{LF.cpp} is a model-dependent process, the sequence of steps defined within \texttt{CollectLoopFunctions.wl} is highly general and extensible and it should be easy to adopt them for higher order results.

\subsection{Repeated index summations}

The parameter set of the SM involves matrices such as the Yukawa couplings, where the indices describe fermion flavour. BSM scenarios can have additional matrices, or even masses with flavour indices (e.g. sfermions in the MSSM). The matched expressions for a Wilson coefficient can involve complicated products of such flavourful parameters, with a variety of repeated indices and as many free indices as there are fermions in the corresponding operator. 

Explicitly expanding the flavour-sum within the symbolic expression is not preferable, for the following reasons
\begin{enumerate}
    \item[(i)] The expressions for some coefficients such as ${\cal C}_H$ already contain thousands of terms. Explicit expansion will make the expressions unmanageable at the Mathematica level. 

    \item[(ii)] The flavour indices are not just contracted across masses and matrices (Yukawa-like couplings). 
    Indices can appear within the masses inside loop-function calls.
\end{enumerate}
An additional complication arises from the fact that the number of repeated indices is not fixed across individual terms of any expression: we cannot do the sum numerically by using nested loops on the C++ side, because we will not know \textit{a priori} how deep the nesting will be for each case. Also, such an approach is not easily generalizable. In \OpToCpp, we circumvent these issues as follows: 
\begin{enumerate}
    \item[(i)] For each term in the matched expression for a coefficient, we wrap the products of (flavour) tensor objects into an \texttt{EinsSum} function call, which is declared and defined in \texttt{OperatorImport.h(.cpp)}.
    
    \item[(ii)] \texttt{EinsSum} can deal with different tensor objects, whether they are masses (1d vectors), Yukawa couplings (matrices), or loop-function calls containing flavourful masses as arguments. This is achieved by using the modern C++ feature of sum types, more specifically the \texttt{std::variant} class template. The function declaration for \texttt{EinsSum} appears as follows,
    \begin{quote}
    \texttt{std::complex<double> EinsSum(\\
       \hspace*{0.5cm} std::vector<std::variant<a, b, c> > tensor\_objs, \\
       \hspace*{0.5cm} std::vector<std::vector<int> > index\_order,\\
       \hspace*{0.5cm} std::vector<int> free\_indices\\
    );}
    \end{quote}
    It thus keeps track of not only the order of the indices (repeated as well as free) within each object but also the specific free indices that appear in a given term. Here, (\texttt{a}, \texttt{b}, \texttt{c}) represent possible data types that can describe objects with indices appearing in the matched expressions. 

    \item[(iii)] Arbitrarily nested loops are avoided by a clever use of the number of flavours ($n$), which in the current version is restricted to $n=3$, and the number of repeated indices ($m$) to first generate a single vector containing all $n^m$ combinations of index-values and then iterate over this vector just once to evaluate the sum. The \texttt{index\_order} argument supplied to \texttt{EinsSum} is used to appropriately substitute the index values. 
    
    This approach not only avoids nested loops but is also generalizable to any number of repeated indices and any number of flavours.
\end{enumerate}

\subsection{Expression translation}

\begin{figure}[h]
\begin{lstlisting}[style=cpp, caption={Before and after of an expression with non-trivial index contractions.}, label={lst:expr-translation}]

// expression within the Matchete results

hbar * (Coupling[c\[Gamma], {}, 0]^6 * Coupling[\[Mu]t, {}, 0]^6
     * Bar[Coupling[cHqu, {Index[d$$1, Flavor], Index[d$$2, Flavor]}, 0]] 
     * Bar[Coupling[cHqu, {Index[d$$3, Flavor], Index[d$$4, Flavor]}, 0]]
     * Bar[Coupling[cHqu, {Index[d$$5, Flavor], Index[d$$6, Flavor]}, 0]]
     * Coupling[cHqu, {Index[d$$1, Flavor], Index[d$$4, Flavor]}, 0]
     * Coupling[cHqu, {Index[d$$3, Flavor], Index[d$$6, Flavor]}, 0]
     * Coupling[cHqu, {Index[d$$5, Flavor], Index[d$$2, Flavor]}, 0]
     * LF[{Coupling[mqt, {Index[d$$1, Flavor]}, 0], 
          Coupling[mqt, {Index[d$$3, Flavor]}, 0],
          Coupling[mqt, {Index[d$$5, Flavor]}, 0],
          Coupling[mut, {Index[d$$2, Flavor]}, 0],
          Coupling[mut, {Index[d$$4, Flavor]}, 0],
          Coupling[mut, {Index[d$$6, Flavor]}, 0]}, 
          {1, 1, 1, 1, 1, 1, 0}]) / Coupling[s\[Gamma], {}, 0]^6


// expression after exporting to C++

(hbar*pow(cgamma,6)*pow(muTilde,6))/pow(1-pow(cgamma,2),3)
    *EinsSum({LoopFunc({mqt,mqt,mqt,mut,mut,mut},132,mubarsq),
              Yu,Yu,Yu,Yu_c,Yu_c,Yu_c}, {{1,3,5,2,4,6},
              {1,4},{3,6},{5,2},{1,2},{3,4},{5,6}},{})
\end{lstlisting}
\end{figure}
The translation from the Mathematica syntax, used by \matchete, to C++ syntax involves systematic manipulation of terms and a reorganization of variables which do/do not have flavour indices. Listing~\ref{lst:expr-translation} illustrates how the matched expressions computed by \matchete look like after having been exported to C++. 
One can notice that:
\begin{itemize}
    \item Variable names are suitably modified, in accordance with the replacement rules defined in 
    \textsf{Cell~[6]} of Listing~\ref{lst:OpExp.nb}.

    \item Exponentiation within the Mathematica expression is converted to a call to the \texttt{pow} function in C++.

    \item \texttt{Bar[Coupling[x,\_,\_]]} is replaced by \texttt{Coupling["x\_c",\_,\_]} and then interpreted as the variable \texttt{x\_c}, i.e. the complex/Hermitian conjugate of \texttt{x} which is defined and initialized internally.

    \item All parameters with non-zero flavour dimensions, are collected within the \texttt{EinsSum} function call, while keeping track of the order in which the repeated indices appear. For instance, the loop-function contains the repeated indices in the order (\texttt{d\$\$1}, \texttt{d\$\$3}, \texttt{d\$\$5}, \texttt{d\$\$2}, \texttt{d\$\$4}, \texttt{d\$\$6}). This ordering is reflected in the first element of the second argument to \texttt{EinsSum}, i.e. \texttt{\{1,3,5,2,4,6\}}.

    \item The \matchete \texttt{LF} function call is translated to a call to the \texttt{LoopFunc} function object in C++. While, the order of masses is preserved, the loop function structure (\texttt{\{1, 1, 1, 1, 1, 1, 0\}}) is replaced by a single integer ($132$).
    Additionally, the matching scale \texttt{mubarsq} is added as an extra function argument.
\end{itemize}
\subsection{Optimizations}

Once the matched expressions are converted to C++ functions, with proper incorporation of loop-functions and repeated index summations, each Wilson coefficient method call simply amounts to a sequence of arithmetic operations. Then, what affects the speed of execution is the number of terms in the expression for a particular Wilson coefficient. Coefficients such as ${\cal C}_H$, that involve a large number of operations, take longer to evaluate in comparison to coefficients like ${\cal C}_G$, which evaluate almost instantly. It is important to note that the complexity of a given Wilson coefficient can vary based on the choice of the UV model. We present some benchmarks highlighting this in Section~\ref{sec:benchmarking}. In the current version of \OpToCpp, we have adopted optimizations at 2 different levels:
\begin{enumerate}
    \item Within individual terms of the matched expressions:
    \begin{itemize}
        \item The \texttt{EinsSum} function (defined in \texttt{OperatorImport.cpp}) appears repeatedly in the matched expression. This function is used to sum over repeated flavour indices within terms containing products of masses, matrices and loop-functions. Each call involves an iteration over a $3^m$-element vector ($m$ being the number of repeated indices). We noticed significant speed up by introducing parallelisation through OpenMP. For instance, a single call to the most complex method, i.e., \texttt{cH()} which used to take >4s was brought down to $\sim$1.8s after parallelisation (on an M3 Macbook Pro with 8GB RAM).\footnote{The relatively slow execution time of 1.8s for \texttt{cH()} even after parallelisation is due to the highly complex form of the matching relations for the MSSM, with >5000 terms in the expression for \texttt{cH()}. For a simpler model, such as a Singlet Scalar Extension of the SM \cite{Haisch:2020ahr}, the same coefficient evaluates in around $10^{-5}$s.}
    \end{itemize}

    \item When dealing with large numbers of Wilson coefficients, we have implemented optimizations at the C++ back-end as well as at the level of the Python functions: 
    \begin{itemize}
        \item The \texttt{batch\_eval} (defined as a method inside \texttt{<model>.cpp}) was introduced to process large numbers of Wilson coefficient calls, $O(100) - O(1000)$ within a single call. This ensures that for large number of evaluations the Python -- C++ boundary is not crossed repeatedly, thus eliminating the minor overheads that may accumulate over 1000 function calls. 
        With parallelisation (once again with OpenMP), it can be ensured that tasks, differing by orders of magnitude with respect to the execution times (for instance, ${\cal C}_H$ versus ${\cal C}_G$), are appropriately distributed across threads. This way a few expensive function calls do not hinder the fast execution of (relatively) lighter functions.

        \item With the \texttt{ThreadPoolExecutor} class in Python: For creating dataframes individual rows can be built independently, so we implemented parallelisation on the Python side through the \texttt{ThreadPoolExecutor} class which is provided by the \texttt{concurrent.futures} module.  
    \end{itemize} 
\end{enumerate}

While the parallelisation within \texttt{EinsSum} is always there and it adds \texttt{libomp} as a dependency, the optimizations introduced for batch processing are optional. As a consequence, Python functions defined in the \texttt{utils.io} module that deal with large numbers of  coefficient evaluations, e.g. \texttt{write\_to\_yaml}, \texttt{write\_to\_wcxf} provide the user with the option to specify the mode of execution, through the argument, \texttt{opt = "seq" | "par"}. The performance of each option for various cases is discussed in Sec~\ref{sec:benchmarking}.

\subsection{Interfacing with different data formats}

We have bundled simple functions within the \texttt{utils} module that allow us to interface and interact with the following data formats  (see Appendix~\ref{app:doc} for details on the specific functions).
\begin{enumerate}
    \item[(i)] \texttt{WCxf} -- The \texttt{PyYAML} library is used to not only define functions that read parameter values from an input file but also to write the values of the computed Wilson coefficients to a \texttt{.yaml} file in the Wilson Coefficient exchange format (\texttt{WCxf}). 
    
    An additional function is provided to write the values of specific model parameters and evaluated coefficients
    to a \texttt{.yaml} file, while retaining the \matchete nomenclature. 

    \item[(ii)] \texttt{Pandas} and \texttt{CSV} -- We provide functions to enable the creation of dataframes for storing the results of multi-parameter scans. For writing the output to 
    \texttt{.csv} files, we recommend using \texttt{to\_csv} function provided by the \texttt{Pandas} library. 
\end{enumerate}

\section{Benchmarking and performance}\label{sec:benchmarking}

\begin{table}[!htb]
    \centering
    \renewcommand{\arraystretch}{1.4}
    \begin{tabular}{|c|c|c|c|}
        \hline
        \multicolumn{4}{|c|}{\textbf{Execution times for the \texttt{WCxf} writer}}\\
        \hline
        
        UV Model&
        Mode&
        M3, 8GB&
        i7, 16GB\\
        \hline

        \multirow{2}{*}{MSSM}&
        Sequential, all WCs&
        30s&
        11s\\

        &
        Parallel, all WCs&
        12s&
        5s\\
        \hline

        \multirow{2}{*}{Type I Seesaw}&
        Sequential, all WCs&
        0.5s&
        0.3s\\

        &
        Parallel, all WCs&
        0.2s&
        0.2s\\
        \hline
        
        \multirow{2}{*}{SM + Singlet Scalar}&
        Sequential, all WCs&
        0.1s&
        0.1s\\

        &
        Parallel, all WCs&
        0.1s&
        0.1s\\
        \hline
        \hline
        \multicolumn{4}{|c|}{\textbf{Execution times for the dataframe builder}}\\
        \hline
        
        UV Model&
        Grid Shape&
        M3, 8GB&
        i7, 16GB\\
        \hline

        MSSM&
        500 points, 20 WCs&
        560s&
        400s\\

        \hline

        Type I Seesaw&
        500 points, 20 WCs&
        10s&
        7s\\
        \hline

        SM + Singlet Scalar&
        500 points, 20 WCs&
        1s&
        0.6s\\

        \hline
        \hline
    \end{tabular}
    \caption{A summary of the execution times for the (i)~\texttt{write\_to\_wcxf} and (ii)~\texttt{create\_dataframe} functions, for different UV models, computational setup, and the mode of operation (for the former). }
    \label{tab:performance}
\end{table}

To illustrate the performance of our code, Table~\ref{tab:performance} lists the execution times for (i) the \texttt{WCxf} writer function, \texttt{write\_to\_wcxf} and (ii) the dataframe builder,  \texttt{create\_dataframe} for the MSSM, the Type I Seesaw model and the Singlet Scalar extension of the SM. 
The performance was recorded on an M3 Macbook Pro with 8 GB RAM and an i7-9700 Ubuntu desktop with 16 GB RAM. For the \texttt{WCxf} writer, the performance is compared for single-threaded, sequential execution with multithreaded execution. For the dataframe builder, the same set of 20 Wilson coefficients was evaluated for each UV model.  We notice that
\begin{enumerate}
    \item The amount of available RAM has a pronounced effect on the execution time. Even the older Intel processor performs noticeably better than the relatively modern Apple~M3.
    
    \item The complexity of the UV Model informs the complexity of the matching relations, i.e. the expressions for individual Wilson coefficient methods. We see that for a simpler model such as, the Singlet-Scalar extension of the SM \cite{Haisch:2020ahr}, the sequential execution of \texttt{write\_to\_wcxf} improves by 2 orders of magnitude. 
\end{enumerate}

The performance of \texttt{write\_to\_wcxf} and \texttt{create\_dataframe} should not be compared directly, as these functions execute very different operations. The former evaluates all $O(10^3)$ SMEFT Warsaw basis coefficients once, while the latter evaluates the specified coefficients for multiple parameter combinations. These functions have been optimised in different ways. Multithreading is implemented for \texttt{write\_to\_wcxf} through the \texttt{batch\_eval} method defined within the C++ model class, which benefits from parallelisation through OpenMP. On the other hand, for \texttt{create\_dataframe}, parallel row-creation is implemented using a thread-pool executor in Python. The parallelisation within the \texttt{EinsSum} function (which gets executed multiple times even within a single coefficient) is present in all cases.

\section{Conclusions}\label{sec:conclusion}

The matching of complex UV models onto lower-energy EFTs, nowadays helped by automated tools like \matchete, can result in highly complicated analytical expressions.  There might be (i) 100s\,--\,1000s of individual terms in the matched expressions of some Wilson coefficients, (ii) a wide variety of loop-functions encapsulating integrals involving multiple heavy masses and (iii) variable patterns of repeated index contractions among parameters carrying flavor indices. To be able to perform numerical analyses, e.g.\ for global fits, it is then necessary to translate the analytical expressions into efficient, runnable code. This task is performed by \OpToCpp, a novel hybrid Mathematica and C++ tool presented in this paper.  

Concretely, \OpToCpp takes the EFT model file, which defines the operator basis and Wilson coefficients, together with the file storing the UV-to-EFT matching relations from \matchete and converts them into a C++ class and its methods. 
This can be used to build a C++ executable or a Python library for performing numerical analyses. The workflow consists of three main steps. First, the EFT information is parsed into a JSON file. Second, the unique loop functions and their degeneracy limits are translated into C++ friendly form. Finally, the matching results  are systematically converted and exported to C++. 

These steps are performed through the Mathematica component of \OpToCpp and result in a set of C++ header and source files.  
Having access to the Wilson coefficients as C++ functions makes it possible to directly perform numerical evaluations, or use them in other tools, such as \texttt{HEPfit}, for numerical analyses. 
One can also create a shared object (.so) library for usage with Python. To connect with the wider ecosystem of Python libraries (e.g.\ \texttt{Wilson}, \texttt{SMEFiT}) and data formats (such as \texttt{WCxf}) used in EFT analyses, a Python-based front end is provided with \OpToCpp. 

In addition to describing the technical design and documenting the various Mathematica and Python functions of interest to the user, this paper provides installation instructions and usage examples for the tool. \OpToCpp is a free and open-source project distributed under the GNU General Public License, GPLv3. The source code as well as the examples discussed in this paper are publicly available on \href{https://github.com/BSM-EFT/OperatorToCpp}{Github}. 

The version described here is released as v1.0~\cite{operator-to-cpp:github}. Plans for future versions concern, for instance, the inclusion of higher-tensor coefficients in flavour space and a more rigorous treatment of degeneracies in loop-functions. Moreover, interfaces to other tools are envisaged, thus enlarging the overall EFT analysis ecosystem \cite{Proceedings:2019rnh,Aebischer:2023nnv}. The generic and modular nature of the codebase should make it convenient to apply it to matching results involving dimension-8 SMEFT operators (for simpler UV models), two-loop matching (once such results become available) \cite{Fuentes-Martin:2024agf}, and EFTs involving BSM degrees of freedom \cite{Anisha:2019nzx,Banerjee:2020jun,Dermisek:2024ohe}.

\section*{Acknowledgements}
This work was supported in part by the French ANR under project ANR-22-CE31-0022-02 (EFTatLHC). 
AL is supported by FAPESP under grants no.~2018/25225-9 and 2021/01089-1 and CNPq grant no.~300278/2025-0.
SP is supported by MCIU/AEI/10.13039/501100011033 (grants CEX2023-001292-S and PID2023-146220NB-I00).
FW~acknowledges support by the Deutsche Forschungsgemeinschaft (DFG, German Research Foundation) under grant 396021762 – TRR~257: \textit{Particle Physics Phenomenology after the Higgs Discovery}. 

\appendix

\section{Documentation of some useful functions}\label{app:doc}

\subsection{Mathematica functions}
In the Mathematica user interface, the following functions are available to the user to process the matched results for export and for generating C++ and Python files.
\begin{itemize}
    \item \texttt{\color{codekw}ReadEFTInfo[{\color{teal} LowEnergyModel}, {\color{teal} EFT}]}
    
    \textbf{Arguments}: 
    \begin{itemize}
        \item \texttt{{\color{teal} LowEnergyModel}}: A string describing the low energy model. If the model is already present in the \matchete models-database, then simply specifying the name is sufficient. Alternatively, the user can specify the path of a \texttt{.m} file that defines the low-energy model in \matchete format.
        
        \item \texttt{{\color{teal} EFT}}: A string describing the Effective Field Theory and the operator basis. Once again, either the name (if present in the models database) or the path to a \texttt{.m} file can be specified. 
    \end{itemize}
    
    \textbf{Description}: Reads the coupling/coefficient information for the specified EFT and returns an association with the Wilson coefficients as keys and the values specify relevant attributes, e.g.,\\
    \texttt{<|\dots \\ 
    \scalebox{0.8}{"cG" $\rightarrow$ <| "Nf" $\rightarrow$ 0, "SelfConjugate" $\rightarrow$ True, "Permutations" $\rightarrow$ \{ \}  |>,}\\
    \scalebox{0.8}{"cHd" $\rightarrow$ <| "Nf" $\rightarrow$ 2, "SelfConjugate" $\rightarrow$ False, "Permutations" $\rightarrow$ \{ \{1,2\}\}  |>,}\\
    \dots\\
    |>}
    \item \texttt{\color{codekw}ExportWCInfo[{\color{teal} EFT}, {\color{teal} Basis}, {\color{teal} WCInfo}]}
    
    \textbf{Arguments}: 
    \begin{itemize}
        \item \texttt{{\color{teal} EFT}}: A string specifying the name of the specific EFT. 
        \item \texttt{{\color{teal} Basis}}: A string specifying the specific basis, in the context of the EFT.
        \item \texttt{{\color{teal} WCInfo}}: An association, produced by the \texttt{{\color{codekw} ReadEFTInfo}} function,  containing Wilson coefficient metadata. 
    \end{itemize}
    
    \textbf{Description}: Given the specific EFT and Basis names, and Wilson coefficient metadata. It stores the information in the file \texttt{py/wcInfo/<EFT>-<Basis>.json} for later use at the Python end.
    
    \newpage
    \item \texttt{\color{codekw}SimplifyOutput[{\color{teal} matchedResult}, {\color{teal} WCInfo}]}
    
    \textbf{Arguments}: 
    \begin{itemize}
        \item \texttt{\color{teal} matchedResult}: A dictionary storing \matchete output. This typically contains one-loop corrections to the SM couplings, in addition to the matching conditions for the Wilson coefficients.

        \item \texttt{\color{teal} WCInfo}: An association, produced by the \texttt{{\color{codekw} ReadEFTInfo}} function,  containing Wilson coefficient metadata.
    \end{itemize}
    
    \textbf{Description}: Creates a dictionary containing only Wilson coefficients and the corresponding matching conditions, while unwraping the \matchete API and converting the expressions to a lighter intermediate form.

    \item \texttt{\color{codekw}BuildFiles[{\color{teal}ModelName}, {\color{teal}SimplifiedOutput}, {\color{teal}ComplexParams}, {\color{teal}WCInfo}]}

    \textbf{Arguments}:
    \begin{itemize}
        \item \texttt{\color{teal} ModelName}: The name of the UV model.
        \item \texttt{\color{teal} SimplifiedOutput}: Intermediate form of the matching results generated by \texttt{\color{codekw}SimplifyOutput} 
        \item \texttt{\color{teal} ComplexParams}: List of complex-valued parameters.

        \item 
    \texttt{\color{teal} WCInfo}: An association, produced by the \texttt{{\color{codekw} ReadEFTInfo}} function,  containing Wilson coefficient metadata
    \end{itemize}
    
    \textbf{Description}: Creates C++ header and source files  \texttt{ModelName.h/.cpp}, with implementations for each Wilson coefficient method corresponding to the \texttt{ModelName} class. The method bodies are built by manipulating the \texttt{\color{teal}SimplifiedOutput}. Also generates class and method declarations for use at the Python front-end and stores them in the file \texttt{match\_to\_py.pyi}, and bundles the necessary \texttt{pybind11} code inside \texttt{pyBindings.cpp} to enable the creation of a Python module. 

\end{itemize}

\subsection{Python}

The following helper functions are provided through the \texttt{utils} module for use at the Python front-end.

\begin{itemize}
    \item \texttt{\color{codekw}eval\_wc({\color{teal}model},{\color{teal}wc\_name})}
    
    \textbf{Arguments}: 
    \begin{itemize}
        \item \texttt{\color{teal}model}: An instance of the UV model class.

        \item \texttt{\color{teal}wc\_name}: Name of the Wilson coefficient in the \texttt{"cXY\_ij.."} format. \\
        Here $i,j,\dots\in\{1,2,3\}$ denote flavour indices.
    \end{itemize}
    
    \textbf{Description}: Evaluates the specified Wilson coefficient method of a particular instance of the model class. Returns a complex-valued result. 

    \item \texttt{\color{codekw}read\_param\_values({\color{teal}filename})}
    
    \textbf{Arguments}: 
    \begin{itemize}
        \item \texttt{\color{teal}filename}: A string specifying the path of a \texttt{.yaml} file containing values of model parameters.
    \end{itemize}
    
    \textbf{Description}: Reads the contents of a \texttt{.yaml} file and returns a tuple containing (i) a dictionary of (name, value) pairs for fixed parameters and (ii) a dictionary of (name, range) pairs for parameters that vary within a range specified in the \texttt{[min, max, num]} format in the input file.

    \item \texttt{\color{codekw}read\_wc\_names({\color{teal}filename})}
    
    \textbf{Arguments}: 
    \begin{itemize}
        \item \texttt{\color{teal}filename}: A string specifying the path of a \texttt{.txt} file containing the names of Wilson coefficients.
    \end{itemize}
    
    \textbf{Description}: Reads the names of Wilson coefficient names from a \texttt{.txt} file and stores them in a list.

    \item \texttt{\color{codekw}create\_dataframe({\color{teal}model}, {\color{teal}fixed\_pars}, {\color{teal}ranges\_dict}, {\color{teal}wc\_names}, \\ \hspace*{3.4cm} {\color{teal}grid},  {\color{teal}max\_workers},  {\color{teal}**kw})}
    
    \textbf{Arguments}: 
    \begin{itemize}
        \item \texttt{\color{teal}model}: The UV model class.

        \item \texttt{\color{teal}fixed\_pars}: A dictionary containing (name, value) pairs for fixed model parameters.
        
        \item \texttt{\color{teal}ranges\_dict}: A dictionary containing (name, range) pairs for model parameters that
        vary over specific ranges.

        \item \texttt{\color{teal}wc\_names}: A list of Wilson coefficient names.

        \item \texttt{\color{teal}grid} = \texttt{True} (default) | \texttt{False}: A flag for specifying whether the combinations should be created as Cartesian products (the default option) or by simply combining the corresponding entries of each array in \texttt{\color{teal}ranges\_dict}. 

        \item \texttt{\color{teal}max\_workers}: Number of cores over which the multi-threaded task will be distributed. The default value is set to the maximum number or logical cores available. 

        \item \texttt{\color{teal}**kw}: Keyword arguments to specify the renormalization scale and the order of the matching results.
    \end{itemize}
    
    \textbf{Description}: Generates combinations for parameters that vary over specified ranges, evaluates the
    corresponding Wilson coefficients (for the UV model) and returns a \texttt{Pandas} dataframe object containing the parameter and Wilson coefficient names as columns and their values as rows. 

    \item \texttt{\color{codekw}write\_to\_wcxf({\color{teal}filename}, {\color{teal}model}, {\color{teal}eft\_info}, {\color{teal}wc\_names}, {\color{teal}opt})}
    
    \textbf{Arguments}: 
    \begin{itemize}
        \item \texttt{\color{teal}filename}: A string specifying the path of the output \texttt{.yaml} file.

        \item \texttt{\color{teal}model}: An instance of the UV model class.
        
        \item \texttt{\color{teal}eft\_info} = A dictionary specifying the "eft" and the "basis".
        
        \item \texttt{\color{teal}wc\_names} = \texttt{list} | \texttt{None} (default): A list of Wilson coefficient names in the \texttt{WCxf} convention. The default option evaluates the entire list of coefficients in the specified EFT-basis.  

        \item \texttt{\color{teal}opt} = \texttt{"seq"} | \texttt{"par"} (default): An option to specify whether the operation should proceed sequentially, or in parallel either with OpenMP at the C++ backend.
    \end{itemize}
    
    \textbf{Description}: Writes the values of evaluated Wilson coefficients to a \texttt{.yaml} file in the \texttt{WCxf} convention. Internally, we define a variable named \texttt{zero\_cutoff}, with a default value of $10^{-20}$. Any numerical output with an absolute value smaller than this cutoff is treated as zero, and hence omitted from the generated \texttt{wcxf} file.

    \item \texttt{\color{codekw}write\_to\_yaml({\color{teal}filename}, {\color{teal}model}, {\color{teal}param\_dict}, {\color{teal}keys}, {\color{teal}opt})}
    
    \textbf{Arguments}: 
    \begin{itemize}
        \item \texttt{\color{teal}filename}: A string specifying the path of the output \texttt{.yaml} file.

        \item \texttt{\color{teal}model}: An instance of the UV model class.
        
        \item \texttt{\color{teal}param\_dict}: A dictionary containing (name, value) pairs of model parameters.

        \item \texttt{\color{teal}keys}: A list containing a list of model parameter names and a list of Wilson coefficient names which will constitute the keys in the output file.

        \item \texttt{\color{teal}opt} = \texttt{"seq"} | \texttt{"par"} (default): An option to specify whether the operation should proceed sequentially, or in parallel either with OpenMP at the C++ backend.

    \end{itemize}
    
    \textbf{Description}: Writes the values of specified model parameters and evaluated Wilson coefficients to a \texttt{.yaml} file, in the \matchete convention. Again, values of Wilson coefficient below $10^{-20}$ are set to zero by default.

\end{itemize}

\section{C++ user interface}\label{app:code-listings}

\begin{figure}[!htb]
\begin{lstlisting}[style=cpp, caption={\small{A snippet of the file \texttt{MSSM.h} generated by \texttt{OpExp\_MSSM.nb}.}}, label={lst:model.h},]
#pragma once
#include <vector>
#include <string>
#include <complex>
#include <unordered_map>
#include <map>
#include <functional>

const double pi = 3.14159265;

struct Task {
    std::string name;
    std::function<std::complex<double>()> work;
};
 
class MSSM {
  private:
    double hbar = 1/(16 * pow(pi,2));
    double mubarsq = 1000.0;    
    std::complex<double> g1 = 0.0;
    std::complex<double> mPhi = 0.0;
    ...    
    std::vector<std::complex<double>> mut = {0.0, 0.0, 0.0};
    ...
    std::vector<std::vector<std::complex<double>>> Yu = {
            {0.0, 0.0, 0.0}, {0.0, 0.0, 0.0}, {0.0, 0.0, 0.0}};
    std::vector<std::vector<std::complex<double>>> Yuc = {
            {0.0, 0.0, 0.0}, {0.0, 0.0, 0.0}, {0.0, 0.0, 0.0}};

  public:
    MSSM() = default;
    MSSM(std::unordered_map<std::string, std::complex<double>> params, 
         double scale, bool loop);
    
    void updateParams(
            std::unordered_map<std::string, std::complex<double>> params);
            
    double getScale();
    void setScale(double scale);
    void loopContributions(bool loop);
    std::unordered_map<std::string, std::complex<double> > getParams();

    std::complex<double> cllHH(int i1, int i2, double mubarsq, double hbar);
    std::complex<double> cG(double mubarsq, double hbar);
    std::complex<double> ceH(int i1, int i2, double mubarsq, double hbar);
    std::complex<double> cll(int i1, int i2, int i3, int i4, 
                             double mubarsq, double hbar);
    ...
    std::map<std::string, std::complex<double>> batch_eval(
                                const std::vector<Task>& tasks); 
}
\end{lstlisting}
\end{figure}
Listing~\ref{lst:model.h} shows a part of \texttt{MSSM.h}, i.e. the model header file, generated in the final step of Listing~\ref{lst:OpExp.nb}. It shows the MSSM class, its private member variables, i.e., the model parameters, initialized to zero values, and the scale (\texttt{mubarsq}) and loop factor (\texttt{hbar}) initialized with non-zero values. It shows declarations of the default and overloaded constructors. The latter admits a map containing (name, value) pairs for parameters, a double-valued renormalization scale, and a boolean to specify the inclusion (or exclusion) of loop contributions. The \texttt{updateParams()} method enables updating one or more parameter values using a map, while \texttt{getParams()} can be used to examine the values of all parameters at a given time. The renormalization scale can be checked and updated using \texttt{getScale()} and \texttt{setScale(scale)} respectively, whereas \texttt{loopContributions(true|false)} can be used to turn the loop contributions on or off. The method declarations also include Wilson coefficients for SMEFT operators of dimensions 5 and 6.

Additionally, the header file defines the \texttt{Task} struct and the \texttt{batch\_eval(tasks)} method to enable parallel evaluation of a large number of Wilson coefficients. These are only called internally within the Python functions defined in the \texttt{utils} module.

The header file \texttt{MSSM.h} can be imported within a C++ source file, and one can then initialize an instance of the model and evaluate the Wilson coefficient methods. Listing~\ref{lst:main.cpp} shows a simple C++ program that demonstrates the necessary steps. This is essentially the C++ analogue of the first 20 lines of the Python script shown in Listing~\ref{lst:example.py}.

\begin{figure}[!htb]
\begin{lstlisting}[style=cpp, caption={A snippet of a C++ program that creates an instance of the matched model, updates some parameters and prints the values of a few Wilson coefficients.}, label={lst:main.cpp}]
  #include "MSSM.h"
  #include <vector>
  #include <string>
  #include <complex>
  #include <unordered_map>
  #include <iostream>
  
  int main() {

    // create a parameter dictionary
    std::unordered_map<std::string, std::complex<double>> param_dict;
    param_dict.emplace("g1", 0.11);
    param_dict.emplace("m1", 1200);
    param_dict.emplace("mut3", 2000);
    .
    .
    param_dict.emplace("Yu33", 0.9);

    // initialize an instance of the MSSM model with the
    // parameter dictionary, renormalization scale = 1000 GeV 
    // and loop contrbutions turned on
    MSSM sb_model(param_dict, 1000, true); 

    // print the values of a few Wilson coefficients
    std::cout << "cG: {Real = " << cG().real() 
              << ", Imag = " << cG().imag() << "}\n";
    std::cout << "cuB: {Real = " << cuB(2,2).real() 
              << ", Imag = " << cuB(2,2).imag() << "}\n";
    
    return 0;
}
\end{lstlisting}
\end{figure}


\newpage
\bibliography{references.bib}

@misc{operator-to-cpp:github,
  author       = {Prakash, Suraj},
  title        = "{OperatorToC++ release v1.0.0}",
  month        = aug,
  year         = 2026,
  publisher    = {Zenodo},
  howpublished = {\href{https://doi.org/10.5281/zenodo.15599937}{10.5281/zenodo.15599937}},
}

@article{Haisch:2020ahr,
    author = "Haisch, Ulrich and Ruhdorfer, Maximilian and Salvioni, Ennio and Venturini, Elena and Weiler, Andreas",
    title = {Singlet night in Feynman-ville: one-loop matching of a real scalar},
    eprint = {https://arxiv.org/abs/2003.05936},
    archivePrefix = "arXiv",
    primaryClass = "hep-ph",
    reportNumber = "CERN-TH-2020-038, TUM-HEP-1254-20",
    doi = {10.1007/JHEP04(2020)164},
    journal = "JHEP",
    volume = "04",
    pages = "164",
    year = "2020",
    note = "[Erratum: JHEP 07, 066 (2020)]"
}

@article{Zhang:2021jdf,
    author = "Zhang, Di and Zhou, Shun",
    title = "{Complete one-loop matching of the type-I seesaw model onto the Standard Model effective field theory}",
    eprint = {https://arxiv.org/abs/2107.12133},
    archivePrefix = "arXiv",
    primaryClass = "hep-ph",
    doi = "10.1007/JHEP09(2021)163",
    journal = "JHEP",
    volume = "09",
    pages = "163",
    year = "2021"
}

@article{Buchmuller:1985jz,
    author = "Buchmuller, W. and Wyler, D.",
    title = "{Effective Lagrangian Analysis of New Interactions and Flavor Conservation}",
    reportNumber = "CERN-TH-4254/85",
    doi = "10.1016/0550-3213(86)90262-2",
    journal = "Nucl. Phys. B",
    volume = "268",
    pages = "621--653",
    year = "1986"
}

@article{ATLAS:2012yve,
    author = "Aad, Georges and others",
    collaboration = "ATLAS",
    title = "{Observation of a new particle in the search for the Standard Model Higgs boson with the ATLAS detector at the LHC}",
    eprint = {https://arxiv.org/abs/1207.7214},
    archivePrefix = "arXiv",
    primaryClass = "hep-ex",
    reportNumber = "CERN-PH-EP-2012-218",
    doi = "10.1016/j.physletb.2012.08.020",
    journal = "Phys. Lett. B",
    volume = "716",
    pages = "1--29",
    year = "2012"
}

@article{COHERENT:2017ipa,
    author = "Akimov, D. and others",
    collaboration = "COHERENT",
    title = "{Observation of Coherent Elastic Neutrino-Nucleus Scattering}",
    eprint = {https://arxiv.org/abs/1708.01294},
    archivePrefix = "arXiv",
    primaryClass = "nucl-ex",
    doi = "10.1126/science.aao0990",
    journal = "Science",
    volume = "357",
    number = "6356",
    pages = "1123--1126",
    year = "2017"
}

@article{KM3NeT:2025npi,
    author = "Aiello, S. and others",
    collaboration = "KM3NeT",
    title = "{Observation of an ultra-high-energy cosmic neutrino with KM3NeT}",
    doi = "10.1038/s41586-024-08543-1",
    journal = "Nature",
    volume = "638",
    number = "8050",
    pages = "376--382",
    year = "2025",
    note = "[Erratum: Nature 640, E3 (2025)]"
}

@article{Kraml:2025fpv,
    author = "Kraml, Sabine and Lessa, Andre and Prakash, Suraj and Wilsch, Felix",
    title = "{SUSY meets SMEFT: complete one-loop matching of the general MSSM}",
    eprint = {https://arxiv.org/abs/2506.05201},
    archivePrefix = "arXiv",
    primaryClass = "hep-ph",
    reportNumber = "TTK-25-14, P3H-25-036",
    doi = "10.1007/JHEP04(2026)028",
    journal = "JHEP",
    volume = "04",
    pages = "028",
    year = "2026"
}

@article{Fuentes-Martin:2022jrf,
    author = {Fuentes-Mart{\'\i}n, Javier and K{\"o}nig, Matthias and Pag{\`e}s, Julie and Thomsen, Anders Eller and Wilsch, Felix},
    title = "{A proof of concept for matchete: an automated tool for matching effective theories}",
    eprint = {https://arxiv.org/abs/2212.04510},
    archivePrefix = "arXiv",
    primaryClass = "hep-ph",
    reportNumber = "MITP-22-105, TUM-HEP-1443/22, ZU-TH-58/22",
    doi = "10.1140/epjc/s10052-023-11726-1",
    journal = "Eur. Phys. J. C",
    volume = "83",
    number = "7",
    pages = "662",
    year = "2023"
}

@article{Carmona:2021xtq,
    author = "Carmona, Adrian and Lazopoulos, Achilleas and Olgoso, Pablo and Santiago, Jose",
    title = "{Matchmakereft: automated tree-level and one-loop matching}",
    eprint = {https://arxiv.org/abs/2112.10787},
    archivePrefix = "arXiv",
    primaryClass = "hep-ph",
    doi = "10.21468/SciPostPhys.12.6.198",
    journal = "SciPost Phys.",
    volume = "12",
    number = "6",
    pages = "198",
    year = "2022"
}

@article{DasBakshi:2018vni,
    author = "Das Bakshi, Supratim and Chakrabortty, Joydeep and Patra, Sunando Kumar",
    title = "{CoDEx: Wilson coefficient calculator connecting SMEFT to UV theory}",
    eprint = {https://arxiv.org/abs/1808.04403},
    archivePrefix = "arXiv",
    primaryClass = "hep-ph",
    doi = "10.1140/epjc/s10052-018-6444-2",
    journal = "Eur. Phys. J. C",
    volume = "79",
    number = "1",
    pages = "21",
    year = "2019"
}

@article{Giani:2023gfq,
    author = "Giani, Tommaso and Magni, Giacomo and Rojo, Juan",
    title = "{SMEFiT: a flexible toolbox for global interpretations of particle physics data with effective field theories}",
    eprint = {https://arxiv.org/abs/2302.06660},
    archivePrefix = "arXiv",
    primaryClass = "hep-ph",
    reportNumber = "Nikhef-2022-023",
    doi = "10.1140/epjc/s10052-023-11534-7",
    journal = "Eur. Phys. J. C",
    volume = "83",
    number = "5",
    pages = "393",
    year = "2023"
}

@article{Aebischer:2018bkb,
    author = "Aebischer, Jason and Kumar, Jacky and Straub, David M.",
    title = "{Wilson: a Python package for the running and matching of Wilson coefficients above and below the electroweak scale}",
    eprint = {https://arxiv.org/abs/1804.05033},
    archivePrefix = "arXiv",
    primaryClass = "hep-ph",
    doi = "10.1140/epjc/s10052-018-6492-7",
    journal = "Eur. Phys. J. C",
    volume = "78",
    number = "12",
    pages = "1026",
    year = "2018"
}

@article{Aebischer:2017ugx,
    author = "Aebischer, Jason and others",
    title = "{WCxf: an exchange format for Wilson coefficients beyond the Standard Model}",
    eprint = {https://arxiv.org/abs/1712.05298},
    archivePrefix = "arXiv",
    primaryClass = "hep-ph",
    reportNumber = "IFIC-17-61, TUM-HEP-1117-17, LMU-ASC-74-17, IFIC/17-61, KA-TP-38-2017, TUM-HEP-1117/17, LMU-ASC 74/17",
    doi = "10.1016/j.cpc.2018.05.022",
    journal = "Comput. Phys. Commun.",
    volume = "232",
    pages = "71--83",
    year = "2018"
}

@article{DeBlas:2019ehy,
    author = "De Blas, J. and others",
    title = "{$\texttt{HEPfit}$: a code for the combination of indirect and direct constraints on high energy physics models}",
    eprint = {https://arxiv.org/abs/1910.14012},
    archivePrefix = "arXiv",
    primaryClass = "hep-ph",
    reportNumber = "CERN-TH-2019-178, CPHT-RR060.102019, DESY-19-184, DESY 19-184, FTUV/19-1031, FTUV/19-1031,
  IFIC/19-44, KEK-TH-2163, LPT-Orsay-19-36, PSI-PR-19-22, UCI-TR-2019-26, IFIC/19-44",
    doi = "10.1140/epjc/s10052-020-7904-z",
    journal = "Eur. Phys. J. C",
    volume = "80",
    number = "5",
    pages = "456",
    year = "2020"
}

@article{terHoeve:2023pvs,
    author = "ter Hoeve, Jaco and Magni, Giacomo and Rojo, Juan and Rossia, Alejo N. and Vryonidou, Eleni",
    title = "{The automation of SMEFT-assisted constraints on UV-complete models}",
    eprint = {https://arxiv.org/abs/2309.04523},
    archivePrefix = "arXiv",
    primaryClass = "hep-ph",
    reportNumber = "Nikhef 2023-011",
    doi = "10.1007/JHEP01(2024)179",
    journal = "JHEP",
    volume = "01",
    pages = "179",
    year = "2024"
}

@article{Gargalionis:2024jaw,
    author = "Gargalionis, John and Quevillon, Jeremie and Vuong, Pham Ngoc Hoa and You, Tevong",
    title = "{Linear Standard Model extensions in the SMEFT at one loop and Tera-Z}",
    eprint = {https://arxiv.org/abs/2412.01759},
    archivePrefix = "arXiv",
    primaryClass = "hep-ph",
    reportNumber = "DESY-24-184, ADP-24-19/T1258, KCL-PH-TH-2024-72, DESY-24-184; ADP-24-19/T1258; KCL-PH-TH-2024-72;",
    doi = "10.1007/JHEP07(2025)136",
    journal = "JHEP",
    volume = "07",
    pages = "136",
    year = "2025"
}

@article{Belfatto:2025ids,
    author = {Belfatto, Benedetta and Blanke, Monika and Heisig, Jan and Kr{\"a}mer, Michael and Rathmann, Lena and Wilsch, Felix},
    title = "{Toward a comprehensive exploration of flavored dark matter models}",
    eprint = {https://arxiv.org/abs/2511.10490},
    archivePrefix = "arXiv",
    primaryClass = "hep-ph",
    reportNumber = "P3H-25-091, TTP25-043, TTK-25-37",
    doi = "10.1140/epjc/s10052-026-15960-1",
    journal = "Eur. Phys. J. C",
    volume = "86",
    number = "7",
    pages = "802",
    year = "2026"
}

@article{Aebischer:2018iyb,
    author = "Aebischer, Jason and Kumar, Jacky and Stangl, Peter and Straub, David M.",
    title = "{A Global Likelihood for Precision Constraints and Flavour Anomalies}",
    eprint = {https://arxiv.org/abs/1810.07698},
    archivePrefix = "arXiv",
    primaryClass = "hep-ph",
    doi = "10.1140/epjc/s10052-019-6977-z",
    journal = "Eur. Phys. J. C",
    volume = "79",
    number = "6",
    pages = "509",
    year = "2019"
}

@article{Smolkovic:2026cba,
    author = "Smolkovi{\v{c}}, Aleks and Stangl, Peter",
    title = "{Differentiable Multi-scale Effective Field Theory Likelihoods for Beyond the Standard Model Phenomenology}",
    eprint = {https://arxiv.org/abs/2603.15801},
    archivePrefix = "arXiv",
    primaryClass = "hep-ph",
    month = "3",
    year = "2026"
}

@article{CMS:2012qbp,
    author = "Chatrchyan, Serguei and others",
    collaboration = "CMS",
    title = "{Observation of a New Boson at a Mass of 125 GeV with the CMS Experiment at the LHC}",
    eprint = {https://arxiv.org/abs/1207.7235},
    archivePrefix = "arXiv",
    primaryClass = "hep-ex",
    reportNumber = "CMS-HIG-12-028, CERN-PH-EP-2012-220",
    doi = "10.1016/j.physletb.2012.08.021",
    journal = "Phys. Lett. B",
    volume = "716",
    pages = "30--61",
    year = "2012"
}

@article{Grzadkowski:2010es,
    author = "Grzadkowski, B. and Iskrzynski, M. and Misiak, M. and Rosiek, J.",
    title = "{Dimension-Six Terms in the Standard Model Lagrangian}",
    eprint = {https://arxiv.org/abs/1008.4884},
    archivePrefix = "arXiv",
    primaryClass = "hep-ph",
    reportNumber = "IFT-9-2010, TTP10-35",
    doi = "10.1007/JHEP10(2010)085",
    journal = "JHEP",
    volume = "10",
    pages = "085",
    year = "2010"
}

@article{Aebischer:2023nnv,
    author = "Allwicher, Lukas and others",
    editor = "Aebischer, Jason and Fael, Matteo and Fuentes-Mart{\'\i}n, Javier and Thomsen, Anders Eller and Virto, Javier",
    title = {{Computing tools for effective field theories: SMEFT-Tools 2022 Workshop Report, 14{\textendash}16th September 2022, Z{\"u}rich}},
    eprint = {https://arxiv.org/abs/2307.08745},
    archivePrefix = "arXiv",
    primaryClass = "hep-ph",
    doi = "10.1140/epjc/s10052-023-12323-y",
    journal = "Eur. Phys. J. C",
    volume = "84",
    number = "2",
    pages = "170",
    year = "2024"
}

@proceedings{Proceedings:2019rnh,
    author = "Brivio, Ilaria and others",
    editor = "Aebischer, Jason and Fael, Matteo and Lenz, Alexander and Spannowsky, Michael and Virto, Javier",
    title = "{Computing Tools for the SMEFT}",
    eprint = {https://arxiv.org/abs/1910.11003},
    archivePrefix = "arXiv",
    primaryClass = "hep-ph",
    month = "10",
    year = "2019"
}

@article{Fuentes-Martin:2024agf,
    author = "Fuentes-Mart{\'\i}n, Javier and Moreno-S{\'a}nchez, Adri{\'a}n and Palavri{\'c}, Ajdin and Thomsen, Anders Eller",
    title = "{A guide to functional methods beyond one-loop order}",
    eprint = {https://arxiv.org/abs/2412.12270},
    archivePrefix = "arXiv",
    primaryClass = "hep-ph",
    doi = "10.1007/JHEP08(2025)099",
    journal = "JHEP",
    volume = "08",
    pages = "099",
    year = "2025"
}

@article{Fuentes-Martin:2020zaz,
    author = "Fuentes-Martin, Javier and Ruiz-Femenia, Pedro and Vicente, Avelino and Virto, Javier",
    title = "{DsixTools 2.0: The Effective Field Theory Toolkit}",
    eprint = "{https://arxiv.org/abs/2010.16341}",
    archivePrefix = "arXiv",
    primaryClass = "hep-ph",
    reportNumber = "MITP/20-061, IFIC/20-50",
    doi = "10.1140/epjc/s10052-020-08778-y",
    journal = "Eur. Phys. J. C",
    volume = "81",
    number = "2",
    pages = "167",
    year = "2021"
}

@article{Anisha:2019nzx,
    author = "Anisha and Das Bakshi, Supratim and Chakrabortty, Joydeep and Prakash, Suraj",
    title = "{Hilbert Series and Plethystics: Paving the path towards 2HDM- and MLRSM-EFT}",
    eprint = {https://arxiv.org/abs/1905.11047},
    archivePrefix = "arXiv",
    primaryClass = "hep-ph",
    doi = "10.1007/JHEP09(2019)035",
    journal = "JHEP",
    volume = "09",
    pages = "035",
    year = "2019"
}

@article{Banerjee:2020jun,
    author = "Banerjee, Upalaparna and Chakrabortty, Joydeep and Prakash, Suraj and Rahaman, Shakeel Ur and Spannowsky, Michael",
    title = "{Effective Operator Bases for Beyond Standard Model Scenarios: An EFT compendium for discoveries}",
    eprint = {https://arxiv.org/abs/2008.11512},
    archivePrefix = "arXiv",
    primaryClass = "hep-ph",
    reportNumber = "IPPP/20/37",
    doi = "10.1007/JHEP01(2021)028",
    journal = "JHEP",
    volume = "01",
    pages = "028",
    year = "2021"
}

@article{Dermisek:2024ohe,
    author = "Dermisek, Radovan and Hermanek, Keith",
    title = "{Two-Higgs-doublet model effective field theory}",
    eprint = {https://arxiv.org/abs/2405.20511},
    archivePrefix = "arXiv",
    primaryClass = "hep-ph",
    doi = "10.1103/PhysRevD.110.035026",
    journal = "Phys. Rev. D",
    volume = "110",
    number = "3",
    pages = "035026",
    year = "2024"
}


\end{document}